\documentclass[pdflatex,sn-nature]{sn-jnl}

\usepackage{graphicx}%
\usepackage{multirow}%
\usepackage{amsmath,amssymb,amsfonts}%
\usepackage{amsthm}%
\usepackage{mathrsfs}%
\usepackage[title]{appendix}%
\usepackage{xcolor}%
\usepackage{textcomp}%
\usepackage{manyfoot}%
\usepackage{booktabs}%
\usepackage{algorithm}%
\usepackage{algorithmicx}%
\usepackage{algpseudocode}%
\usepackage{listings}%

\begin{document}
\title{Investigation of Super-Poissonian Nonclassical Nature of  Inflaton Field in Flat FRW Universe through Cosmological Mandel's  $\mathbb{Q}$ Parameter}
\author[1]{\fnm{Dhwani} \sur{Gangal}}

\author[1]{\fnm{Sudhava} \sur{Yadav}}

\author*[1]{\fnm{K.K.} \sur{Venkataratnam}}\email{kvkamma.phy@mnit.ac.in}

\affil[1]{\orgdiv{Department of Physics}, \orgname{Malaviya National Institute of Technology Jaipur}, \orgaddress{\street{J. L. N. Marg}, \city{Jaipur}, \postcode{302017}, \country{India}}}

\abstract{This study investigates the nonclassical properties of the inflaton field within the framework of semiclassical gravity by analyzing Cosmological Mandel's $\mathbb{Q}$ parameter for Squeezed Number States (SNS) and Coherent Squeezed Number States (CSNS). Mandel's $\mathbb{Q}$ parameter serves as a critical tool for identifying nonclassical states by differentiating between sub-Poissonian and super-Poissonian statistics. The values of Cosmological Mandel's $\mathbb{Q}$ are positive shows the super-Poissonian non-classical nature of inflaton for Squeezed Number States (SNS) and Coherent Squeezed Number States (CSNS). These results provide deeper insights into the statistical properties of quantum states in early-universe cosmology and emphasize the relevance of SNS and CSNS in understanding quantum effects on cosmic inflation and particle production.}


\maketitle
\section{Introduction}

Cosmology provides a detailed analysis of the origin and evolution of FRW Universe \cite{kubik_origin_2022, moore_big_2014}. In this reference, the Big Bang Theory successfully suggests the best understanding of its structure and expansion of the universe as per the theory of inflation. Still, it has limited justification for the horizon, singularity, homogeneity, flatness, monopole, etc. problems of the universe. It is easy to explain the inflationary period using a slowly-varying massive homogeneous scalar field i.e. inflaton. As proposed by Guth \cite{guth1981cosmological} inflationary theory using an inflaton field can provide a solution to these problems. Out of all available explanations, the simplest explanation is based on the time-dependent exponential expansion (Hubble Rule) of the universe. Where the total energy density of the universe, is dominated by the potential energy density of the scalar field (inflaton) keeping the kinetic contribution much smaller. In cosmology, it is considered that this potential energy of the inflaton field is responsible for negative pressure on space-time generated fast exponentially expansion of universe initially. After inflation, quasi-periodic motion becomes important for the inflaton field with timely decreasing amplitude. This motion of the inflaton field generated various particles in the universe, created re-thermalization in the Universe, and made it hot again \cite{green_cosmological_2022}. Evolution and inflation field at initial stage are responsible for reheating in universe \cite{albrecht_inflation_1994, albrecht_reheating_1982, kofman_reheating_1994, allahverdi_reheating_2010}. This change of phase of the universe shall be optimized using other reheating parameters for a better understanding of standard matter production in the universe \cite{cook_reheating_2015, martin_first_2010, yadav2024reheating, dai_reheating_2014, yadav2024mutated}. 
In general cosmological principles, the behavior of the universe is considered as isotropic and homogeneous. Its Cosmological description can also be produced using Friedmann and Einstein equations of scalar fields. Considering the assumptions, while gravity is considered as classical, Friedmann equations \cite{mohajan_friedmann_2013, suresh_particle_2004} are valid at all stages of the universe. Further, various researchers have shown that quantum fluctuations, density fluctuations, and other quantum properties of matter fields in the universe are also important in cosmology \cite{martin2016quantum}. Semi-Classical Theory of Gravity (SCTG) provides a suitable approach to understanding these quantum effects within classical gravity. SCTG has also demonstrated that gravitational field has a close relation with quantized matter field with curved space-time\cite{kim_one-parameter_1999, finelli_quantum_1999, geralico_novel_2004, padmanabhan_gravity_2005}. Quantum mechanical representation of gravity and matter fields is significant for the best understanding of the Universe, but sometimes for simplicity of cosmological description, in the absence of a suitable quantum theory of gravity, quantum effects were taken negligible at an early stage. Further researchers \cite{gangal2024density, robertson1936kinematics, shaviv2011did, zel1971creation, bergstrom2006cosmology, ellis1999cosmological, kuo_semiclassical_1993, caves_quantum-mechanical_1981, matacz_coherent_1994, suresh_thermal_2001, suresh_squeezed_1998, suresh_nonclassical_2001, lachieze-rey_cosmological_1999,takahashi_thermo_1996, xu_quantum_2007}, has started using SCGT to explain the inflation. The best explanation for this inflationary theory is provided by Friedmann-Robertson-Walker (FRW). Scalar field and FRW metric within Friedmann’s equation of classical gravity show the validity of inflationary scenarios. This shows that semi-classical Friedmann equations with classical consideration of gravity and quantum consideration of matter field(s) can create a genius cosmological model to produce a suitable picture of the early universe. Here it can be easily realized that quantum fluctuations as well as quantized matter fields have an important role in cosmology even the quantum effect of gravity is considered negligible. Within the context of the general theory of relativity, semiclassical Einstein's field equations established the fundamental idea of the Friedmann-Lema{\^ \i}tre-Robertson-Walker metric \cite{mahajan_particle_2008, lachieze-rey_cosmic_1995, ellis_cosmological_1998, carvalho_scalar_2004, gangal2024density}. Recently many researchers have started using quantum consideration of inflaton within SCTG \cite{bak1998quantum, kim1999thermal, gangal2024density}. They had successfully shown that the cosmological inflation depends on the quantum effect of inflaton, thermal condition, re-thermalization, one-loop effective potential \cite{guth1985quantum}, particle production \cite{gangal2024density}  probability distribution \cite{habib1992stochastic}, quantum inflaton in stochastic inflationary scenarios \cite{linde1994big} etc. Nowadays SCTG with other suitable quantum effects become an important tool to explain cosmic evolution for a complete inflation period in a matter-dominated universe. 
Kennard et. al. working in the field of optical physics, for detailed analysis of wave packet quantum mechanically \cite{kennard_zur_1927,venkataratnam_particle_2004} involved the phenomena of quantum squeezed states in optics. Thereafter Squeezed States (SS) and Coherent Squeezed (CSS) became an important class for the analysis of the quantum effect of inflaton in SCTG and cosmology \cite{bakke_geometric_2009, kennard_zur_1927}. Squeezed States (SS) are also receiving much attenuation in cosmology evolution for defining stress tensor, particle production, decoherence phenomena, state polarization, entanglement phenomena, relic graviton, negative energy density, perturbations created, quantum fluctuations as they are nonclassical states with various quantum effects in cosmology. Their quantum number representation for complex and scalar fields is also becoming important for analysis of the expanding and anisotropic nature of FRW universe \cite{gangal2024density, stoica_friedmann-lemaitre-robertson-walker_2016, hu_anisotropy_1978, dhayal_quantum_2020, fischetti_quantum_1979, hartle_quantum_1979, hartle_quantum_1980, hartle_quantum_1981}. Few researchers have described particle creation in expanding universe for oscillating inflaton field \cite{anderson_effects_1983, campos_semiclassical_1994, geralico_novel_2004-1, pedrosa_exact_2007, lopes_gaussian_2009}.Even though there is same proportionality in the power-law expansion of inflaton obtained through quantum and classical laws, but still there is many other discrepancies were reported in the results obtained through classical and semiclassical gravity, like for correction to expansion obtained through semiclassical gravity not showing oscillatory behaviour as obtained in classical gravity, quantum effect on particle creation, possibility distribution of close or open nature of universe \cite{gangal2024density, robertson1936kinematics, shaviv2011did, zel1971creation, bergstrom2006cosmology, ellis1999cosmological, kuo_semiclassical_1993, caves_quantum-mechanical_1981, matacz_coherent_1994}. This show importance and relevance of quantum mechanical behaviour consideration in cosmic inflationary theories \cite{gangal2024density, robertson1936kinematics, shaviv2011did, zel1971creation, bergstrom2006cosmology, ellis1999cosmological, kuo_semiclassical_1993, caves_quantum-mechanical_1981, matacz_coherent_1994, suresh_thermal_2001, suresh_squeezed_1998, suresh_nonclassical_2001, venkataratnam_nonclassical_2010,venkataratnam_density_2008, venkataratnam_oscillatory_2010, venkataratnam_behavior_2013, dhayal_quantum_2020-1, lachieze-rey_cosmological_1999,takahashi_thermo_1996, xu_quantum_2007}, using non-classical state formalism become more important and grabbing much attention to researchers \cite{koh_gravitational_2004, lopes_gaussian_2009-1, lachieze-rey_theoretical_1999, lopes_gaussian_2009-2, sinha_[no_2003, shaviv_did_2011, berger1978classical, berger1981scalar, grishchuk1990squeezed, brandenberger1992entropy, brandenberger1993entropy, kuo1993semiclassical, matacz1993quantum, albrecht1994inflation, gasperini1993quantum, hu1994squeezed}.\\
Hence present paper aims to investigate of massive minimal nonclassical nature of inflaton field in a flat FRW Universe through cosmological Mandel’s $\mathbb{Q}$ Parameter in SNS and CSNS. It is important in the investigation that squeezed states behave like many-particle states so here related fields can be classical, but their statistical properties greatly differ from coherent states whose related fields behave like highly nonclassical.  As quantum-optical phenomena is gaining much attention in quantum cosmology, it is important to determine the statistical nature of various particles in a cosmological model. This can easily be obtained by determining the  Mandel’s $\mathbb{Q}$ parameter. The $\mathbb{Q}$ parameter of Mandel's is helpful for examining nonclassical nature of inflaton field, also a characteristic of deviation in photon statistics from Poisson photon statistics for infalon field, it as an indicator of photon number distribution exhibits oscillatory behavior. Hence we investigated and determined of $\mathbb{Q}$ parameter in the cosmological reference and associated cosmological parameters. As SNS and CSNS states may fluctuate widely in the cosmological background of the early universe.\\
In this study, we descried nonclassical nature of inflaton field in a flat FRW Universe through cosmological Mandel’s $\mathbb{Q}$ Parameter in SNS and CSNS. in the framework of SCGT \cite{nieto1997displaced, ellis1999deviation, penzias1965measurement, handley_curvature_2021}. These states are greatly used in cosmology for explaining the production of particles, entropy enhancement \cite{gasperini1993quantum}, gravitational wave detection \cite{savage1986inhibition} and inflationary scenario \cite{albrecht1982cosmology} etc. The first section of the paper describes the current status in the field of cosmology particularly for nonclassical nature of inflaton field and the importance of the proposed work, second section describes the energy-momentum tensor. Third section talks about the formulation of Squeezed Number State (SNS) for Non-classical FRW metric under SCTG, in forth section describe formulation of Coherent Squeezed Number States (CSNS) for Non-classical FRW metric under SCTG. Fifth section shows the nonclassical nature of inflaton field for flat FRW Universe through Cosmological Mandel’s $\mathbb{Q}$ parameter determined for both SNS as well as CSNS. The sixth section describes the outcome of the research paper. 

\section{Energy-Momentum Tensor in SCTG}

Number of contemporary cosmological models are developed using Einstein's classical gravity field equations with a scalar field utilizing the FRW metric. While establishing a cosmological model, the matter field is described as quantum mechanical, with the background metric as classical known as SCTG. In this case, the gravitational field may be described using semi-classical gravity theory as follows (hence, we assume the unit system c=$\hslash$=1 and {G}=$\frac{1}{\mathit{m}_{\mathit{p}}^2}$).

\begin{equation} \label{2.1}
\mathcal{E}_{\mu \nu }=\frac{8\pi\left\langle :\mathcal{T}_{\mu \nu }:\right\rangle
}{\mathit{m}_{\mathit{p}}^2}.
\end{equation}

Where $\left\langle :\mathcal{T}_{\mu \nu }:\right\rangle$ is the normal ordered expectation values of the energy-momentum tensor and \(\mathcal{E}_{\mu \nu }\) is an Einstein tensor. The time-dependent Schrodinger equation is satisfied by the quantum state $|\psi\rangle$, can be expressed as  

\begin{equation} \label{2.2}
\mathit{i}\frac{ \partial }{\partial \mathit{t}}|\psi\rangle  = \overset{\wedge }{\mathcal{H}}|\psi\rangle,
\end{equation}

where \(\overset{\wedge }{\mathcal{H}}\) represents the Hamiltonian operator. Generalized coordinates \(\left(\mathit{r}_1,\mathit{r}_2,\right. \mathit{r}_3\), \(\mathit{r}_4\)) for FRW space-time follow the relation as

\begin{equation} \label{2.3}
\left(\mathit{d}\mathit{r}_1^2 + \mathit{d}\mathit{r}_2^2 + \mathit{d}\mathit{r}_3^2\right)\mathcal{G}^{2 }(\mathbf{t})-\mathit{d}\mathit{r}_4^2=\mathit{d}\mathit{s}^{2 },
\end{equation}

where $\mathcal{G}(\mathbf{t})$ is known as scale factor. Further Lagrangian density $\mathfrak{L}$ { }for inflaton is 

\begin{equation} \label{2.4}
\mathfrak{L} = -\frac{\sqrt{(-\mathfrak{g})}\left(\mathfrak{g}^{\mu \nu }\partial _{\mu }\Phi  \partial _{\nu }\Phi + m^2\Phi  ^2\right)}{2}.
\end{equation}

Since $\Phi$ is a homogeneous scalar field \cite{gangal2024density, gangal2024particle}. Eq (\ref{2.4}) can be rewritten using metric (\ref{2.3}) as

\begin{equation} \label{2.5}
{\mathfrak{L}} =\frac{{\mathcal{G}}^3(\mathbf{t})(\dot{\Phi}^2-m^2\Phi^2)}{2}.
  \end{equation}
  
K-G (Klein-Gordon) Eq. derived using equation (\ref{2.4}) takes the form as

\begin{equation} \label{2.6}
\ddot{\Phi  }+\frac{3\dot{\mathcal{G}} (\mathbf{t})}{\mathcal{G} (\mathbf{t})}\dot{\Phi  }+ m^2\Phi  =0,
\end{equation}

where the Hubble parameter $\mathfrak{H}$=\(\frac{\dot{\mathcal{G}}(\mathbf{t})}{\mathcal{G}(\mathbf{t})}\) and momentum conjugate of $\overset{\wedge}\Phi$, is $\overset{\wedge}\Pi= \frac{ \partial \mathcal{L}}{\partial \dot\Phi}$. Now on the basis of canonical quantization, the Hamiltonian for inflaton is 

\begin{equation}  \label{2.8}
\langle
:\overset{\wedge }{\mathcal{H}_m}:\rangle=m^2\frac{\mathcal{G}^3(\mathbf{t})\left\langle :\overset{\wedge }\Phi ^2:\right\rangle {}}{2}+\frac{\left\langle :\overset{\wedge }\Pi ^2:\right\rangle {}}{2\mathcal{G}^3 (\mathbf{t})},
\end{equation}

 where $ \left\langle :\overset{\wedge }\Phi ^2:\right\rangle {} $ and $ \left\langle :\overset{\wedge }\Pi ^2:\right\rangle {} $ are normal ordered expectation values. Temporal component of energy-momentum tensor \cite{gangal2024density, gangal2024particle} will be 

\begin{equation} \label{2.9}
\mathcal{T}_{00}= \mathcal{G}^3 (\mathbf{t})\left(m^2\frac{\overset{\wedge }\Phi ^2}{2}+\frac{\dot{\Phi }^2}{2}\right).
\end{equation}

\section{Formulation of Squeezed Number State (SNS) under SCTG}

The definition of a single mode squeezed state is

\begin{equation} \label{3.1}
|\varUpsilon ,\zeta \rangle =\overset{\wedge}W
(\rho ,\Psi ) \mathfrak{D}(\varUpsilon )|0\rangle ,
\end{equation}

where the displacement operator $\mathfrak{D}(\varUpsilon)$ and the squeezing operator \(\overset{\wedge}W(\rho ,\Psi )\) are defined as 

\begin{equation} \label{3.2}
\mathfrak{D}(\Upsilon )=\exp (\Upsilon \overset{\wedge }{\mathit{e}} ^{\dagger }-\Upsilon ^*\overset{\wedge }{\mathit{e}}) ,
\end{equation}

\begin{equation} \label{3.3}
\overset{\wedge}W(\rho ,\Psi )=\exp \frac{\biggr(\rho\exp (-\mathit{i}\Psi )\overset{\wedge }{\mathit{e}}^2-\rho\exp (\mathit{i}\Psi )\overset{\wedge}{\mathit{e}}^ {\dagger2}\biggr) }{2},
\end{equation}

where the squeezing angle ($\Psi $) varies between -$\Pi $ $\leq $ $\Psi $ $\leq $ $\Pi $ while the squeezing parameter ($\rho $) varies between 0 $\leq $ $\rho $ $\leq $ $\infty $. Further \(\overset{\wedge}W(\rho ,\Psi )\) has following properties

\begin{equation} \label{3.4}
\overset{\wedge}W ^{\dagger }\overset{\wedge }{\mathit{e}}\overset{\wedge}W =(\text{cosh$\rho $})(\overset{\wedge }{\mathit{e}})  - (\exp (\mathit{i}\Psi )\text{sinh$\rho $})(\overset{\wedge
}{\mathit{e}} ^{\dagger }),
\end{equation}

\begin{equation} \label{3.5}
\overset{\wedge}W ^{\dagger }\overset{\wedge }{\mathit{e}}^{\dagger }\overset{\wedge}W =(\text{cosh$\rho
$})(\overset{\wedge }{\mathit{e}}^{\dagger }) - (\exp (-\mathit{i}\Psi )\text{sinh$\rho $})(\overset{\wedge }{\mathit{e}}) .
\end{equation}

When squeezing operator is applied to a number state, resulted as following is Squeezed Number State

\begin{equation} \label{3.6}
\overset{\wedge}W(\rho ,\Psi )|n\rangle=|\zeta ,n\rangle.
\end{equation}

The annihilation (\(\overset{\wedge }{\mathit{e}}\)) and creation (\(\overset{\wedge } {\mathit{e}} ^{\dagger }\)) operators have characteristics such as\cite{kim2000nonequilibrium} 

\begin{equation} \label{3.7}
\overset{\wedge }{\mathit{e}}|n,\Phi ,\mathbf{t}\rangle =\sqrt{n}|n-1,\Phi ,\mathbf{t}\rangle, \\
\end{equation}

\begin{equation} \label{3.8}
\overset{\wedge }{\mathit{e}} ^{\dagger }
|n,\Phi ,\mathbf{t}\rangle =\sqrt{n+1}|n+1,\Phi ,\mathbf{t}\rangle,
\end{equation}

\begin{equation} \label{3.10}
\left[\overset{\wedge }{\mathit{e}},\overset{\wedge }{\mathit{e}} ^{\dagger }\right]=1.
\end{equation}

\begin{equation} \label{3.11}
\overset{\wedge }{\mathit{e}} ^{\dagger }\overset{\wedge }{(\mathbf{t})\mathit{e}(\mathbf{t})}|n,\Phi ,\mathbf{t}\rangle =n|n,\Phi ,\mathbf{t}\rangle.
\end{equation}

Hence \(\overset{\wedge } {\mathit{e}} ^{\dagger }\) operated n-times on vacuum state $|0\rangle$ gives rise to $n^{th}$ number state while \(\overset{\wedge }{\mathit{e}}\) operated on $|1\rangle$ produces vacuum state $|0\rangle$.

\begin{equation} \label{3.21}
\frac{[\overset{\wedge }{\mathit{e}} ^{\dagger}]^n |0,\mathbf{t}\rangle}{\sqrt{n!}} =|n,\mathbf{t}\rangle,
\end{equation}

\begin{equation} \label{3.22}
\overset{\wedge }{\mathit{e}}| 0,\mathbf{t} \rangle =|0\rangle, \\
\end{equation}

Here creation and annihilation operators can be computed as \cite{gangal2024density, gangal2024particle}

\begin{equation} \label{3.12}
\overset{\wedge }
{\mathit{e}} ^{\dagger }(\mathbf{t})=\Phi (\mathbf{t})\overset{\wedge }\Pi -\mathcal{G}^3 (\mathbf{t})\dot{\Phi }(\mathbf{t})\overset{\wedge }\Phi.
\end{equation}

\begin{equation} \label{3.13}
\overset{\wedge }{\mathit{e}}(\mathbf{t})=\Phi ^*(\mathbf{t})\overset{\wedge }\Pi -\mathcal{G}^3 (\mathbf{t})\dot{\Phi }^*(\mathbf{t})\overset{\wedge }\Phi, 
\end{equation}

$\overset{\wedge }\Pi$ and $\overset{\wedge }\Phi$ can be computed as using Eqs.(\ref{3.7}-\ref{3.13})

\begin{equation} \label{3.14}
\overset{\wedge }\Phi =\frac{1}{i}\left(\Phi ^*\overset{\wedge }{\mathit{e}} ^{\dagger }-\Phi \overset{\wedge }{\mathit{e}} \right),\\
\end{equation}

\begin{equation} \label{3.15}
\overset{\wedge }\Phi ^2=\left(2\overset{\wedge }{\mathit{e}} ^{\dagger }\overset{\wedge }{\mathit{e}}+1\right)\Phi ^*\Phi -\left(\Phi ^*\overset{\wedge }{\mathit{e}}
^{\dagger }\right)^2-\left(\Phi \overset{\wedge }{\mathit{e}}\right)^2,
\end{equation}

\begin{equation} \label{3.16}
\overset{\wedge }\Pi =\text{i$\mathcal{G}$}^3 (\mathbf{t})\left(\dot{\Phi }\overset{\wedge }{\mathit{e}} -\dot{\Phi }^*\overset{\wedge }{\mathit{e}} ^{\dagger
} \right),\\
\end{equation}

\begin{equation} \label{3.17}
\overset{\wedge }\Pi ^2=\mathcal{G}^3 (\mathbf{t})\left[\left(2\overset{\wedge }{\mathit{e}} ^{\dagger }\overset{\wedge }{\mathit{e}}+1\right)\dot{\Phi }^*\dot{\Phi
}-\left(\dot{\Phi }^*\overset{\wedge }{\mathit{e}} ^{\dagger }\right)^2-\left(\dot{\Phi }\overset{\wedge }{\mathit{e}}\right)^2\right].
\end{equation}

\section{Formulation of Coherent Squeezed Number States (CSNS) under SCTG}

Now we discuss the displacement operation of CSNS. The Single mode coherent state is described as

\begin{equation} \label{4.1}
\mathfrak{D}(\Upsilon )|0\rangle=|\varUpsilon \rangle,
\end{equation}

action of \(\overset{\wedge }{\mathit{e}}\) on $|\varUpsilon \rangle$ is 

\begin{equation} \label{4.2}
\overset{\wedge }{\mathit{e}}|\Upsilon \rangle =\Upsilon |\Upsilon \rangle.
\end{equation}

\(\overset{\wedge }{\mathit{e}}\) and \(\overset{\wedge }{\mathit{e}} ^{\dagger }\) combined with displacement operator $\mathfrak{D}$($\Upsilon $) to satisfy properties as   

\begin{equation} \label{3.18}
\mathfrak{D}^{\dagger }\overset{\wedge }{\mathit{e}} ^{\dagger }\mathfrak{D}=\overset{\wedge }{\mathit{e}} ^{\dagger }+\Upsilon ^*,
\end{equation}

\begin{equation} \label{3.19}
\mathfrak{D}^{\dagger }\overset{\wedge }{\mathit{e}} \mathfrak{D}=\overset{\wedge }{\mathit{e}} +\Upsilon.
\end{equation}

Also operation of $\mathfrak{D}$($\Upsilon $) and $\overset{\wedge}W(\rho ,\Psi )$ on $|0\rangle$ is defined as

\begin{equation} \label{4.5}
\mathfrak{D}(\Upsilon )\overset{\wedge}W(\rho ,\Psi )|0\rangle=|\Upsilon ,\zeta ,0\rangle.
\end{equation}

while operation of $\mathfrak{D}$($\Upsilon $) and $\overset{\wedge}W(\rho ,\Psi )$ on Number State $|n\rangle$ gives the CSNS as

\begin{equation} \label{4.6}
\mathfrak{D}(\Upsilon )\overset{\wedge}W(\rho ,\Psi )|n\rangle=|\Upsilon ,\zeta ,n\rangle.
\end{equation}

Using Eqs. (\ref{3.4}-\ref{3.5}, \ref{3.11}-\ref{3.19}), we get $ \left\langle :\overset{\wedge }\Pi ^2:\right\rangle {} $ for CSNS as \cite{gangal2024density, gangal2024particle}

\begin{align} \label{4.7}
\langle :\overset{\wedge }\Pi^2 :\rangle {}_{\text{CSNS}}=&2\mathcal{G}^6 (\mathbf{t})\biggl[\biggl(n+\frac{1}{2})\sinh^2\rho +n\cosh^2\rho +\frac{1}{2}+\varUpsilon^* \varUpsilon \biggr)\dot{\Phi}^*\dot{\Phi}\nonumber\\
&-\biggl((n+\frac{1}{2})\cosh\rho\sinh\rho-\frac{\varUpsilon^{*2}}{2}\biggr)\dot{\Phi}^{*2}\nonumber\\
&-\biggl((n+\frac{1}{2})\cosh\rho\sinh\rho -\frac{\varUpsilon^2}{2}\biggr)\dot{\Phi}^2\biggr].
\end{align}

using Eqs. (\ref{3.4}-\ref{3.5}, \ref{3.11}-\ref{3.19}), we get $ \left\langle :\overset{\wedge }\Phi ^2:\right\rangle {} $ for CSNS as \cite{gangal2024density, gangal2024particle}

\begin{align} \label{4.10}
\left\langle :\overset{\wedge }\Phi ^2:\right\rangle {}_{\text{CSNS}}=&2\biggr[\biggr(n\text{Cosh}^2\rho +(n+\frac{1}{2})\text{Sinh}^2\rho +\frac{1}{2}+\varUpsilon ^*\varUpsilon\biggr){\Phi }^*{\Phi }\nonumber\\
&-\biggr((n+\frac{1}{2})\text{Sinh$\rho $\text{Cosh$\rho $}}-\frac{\varUpsilon ^*2}{2}\biggr){\Phi }^{*2}\nonumber\\
&-\biggr((n+\frac{1}{2})\text{Sinh$\rho $\text{Cosh$\rho $}}-\frac{\varUpsilon ^2}{2}\biggr){\Phi }^2\biggr].
\end{align}

\section{Nonclassical Nature of  Inflaton Field in Flat FRW Universe through Cosmological Mandel's $\mathbb{Q}$ Parameter}

The nonclassical nature of inflaton field in a flat FRW Universe can be identified through cosmological Mandel’s $\mathbb{Q}$ Parameter. Also, it is important to determine the statistical nature of various particles in a cosmological model, which can easily be obtained by determining Mandel’s $\mathbb{Q}$ parameter. Hence, in this section of the paper, we have computed the $\mathbb{Q}$ parameter in cosmological reference for SNS and CSNS state in semiclassical gravity. In cosmological reference mathematically the Mandel’s $\mathbb{Q}$ Parameter is defined as 

\begin{align} \label{5.1}
\mathbb{Q}=&\frac{<:\hat{\mathcal{N}^2}(\mathbf{t}):>-<:\hat{\mathcal{N}}(\mathbf{t}):>^2-<:\hat{\mathcal{N}}(\mathbf{t}):>}{<:\hat{\mathcal{N}}(\mathbf{t}):>},
\end{align}

Here $<:\hat{\mathcal{N}^2}(\mathbf{t}):>-<:\hat{\mathcal{N}}(\mathbf{t}):>^2=<:(\triangle\hat{\mathcal{N}}(\mathbf{t}))^2:>$  is the dispersion relation for cosmological number operator, using the same in Eqs. (\ref{5.1}), the Mandel’s $\mathbb{Q}$ Parameter can be represented as

\begin{align} \label{5.2}
\mathbb{Q}=&\frac{<:(\triangle\hat{\mathcal{N}}(\mathbf{t}))^2:>-<:\hat{\mathcal{N}}(\mathbf{t}):>}{<:\hat{\mathcal{N}}(\mathbf{t}):>}. 
\end{align}

Now when in Eq. (\ref{5.2}) when $\mathbb{Q}=0$ i.e. $<:(\triangle\hat{\mathcal{N}}(\mathbf{t}))^2:>=<:\hat{\mathcal{N}}(\mathbf{t}):>$ the corresponding state will demonstrate the classical nature otherwise it will show the non-classical cosmological nature \cite{dhayal_quantum_2020-1,dhayal_quantum_2020}. Here also when $(<:(\triangle\hat{\mathcal{N}}(\mathbf{t}))^2:>)<(<:\hat{\mathcal{N}}(\mathbf{t}):>)$ i. e. for negative values of Mandel’s $\mathbb{Q}$ Parameter the behaviour of states is sub-Poissonian non-classical, while for $(<:(\triangle\hat{\mathcal{N}}(\mathbf{t}))^2:>)>(<:\hat{\mathcal{N}}(\mathbf{t}):>)$ i. e. for positive values of Mandel’s $\mathbb{Q}$ Parameter the behaviour of states is super-Poissonian non-classical. For both sub-Poissonian and super-Poissonian non-classical states, mathematical distribution can't be obtained from any classical states. This shows the importance of Mandel’s $\mathbb{Q}$ Parameter to differentiate the classical and non-classical nature of states in cosmology. In Eqs. (\ref{5.2}) $<:\hat{\mathcal{N}}(\mathbf{t}):>$ is known as the normal order expectation value of the cosmological number operator, related to the number of particles produced at any time t concerning the initial time $t_0$ is given by \cite{gangal2024density, gangal2024particle}

\begin{align} \label {5.3}
\mathcal{N}_n\left(\mathbf{t},\mathbf{t}_0\right) &= <n,\Phi ,\mathbf{t}_0\left|\hat{\mathcal{N}}(\mathbf{t})\right|n,\Phi ,\mathbf{t}_0>,
\end{align}

Here n is the quantum mechanical number state, while number operator defined as 
\begin{equation} \label{5.4}
\hat{\mathcal{N}}(\mathbf{t}) = e^{\dagger}e. 
\end{equation}

Here by using Eqs. (\ref{3.11}-\ref{3.12}) in Eq. (\ref{5.4}), normal order expectation value of number operator in cosmological reference can be computed as
 
\begin{equation}\label{5.5}
<:\hat{\mathcal{N}}(\mathbf{t}):> = R^6\dot{\Phi}\dot{\Phi}^*<:\hat{\Phi}^2:>-R^3\dot{\Phi}\Phi^*<:\hat{\Phi}\hat{\Pi}:>-R^3\Phi\dot{\Phi}^*<:\hat{\Pi}\hat{\Phi}:> +\Phi\Phi^*<:\hat{\Pi}^2:>. 
\end{equation}

Now using Eq. (\ref{5.4}) in Eq. (\ref{5.1}), nature of cosmological state can examined using

\begin{align} \label{5.6}
\mathbb{Q}=&\frac{\left\langle :\overset{\wedge}{\mathit{e}}^{\dagger2}{\overset{\wedge}{\mathit{e}}}^2:\right\rangle-\left\langle :\overset{\wedge}{\mathit{e}}^{\dagger}{\overset{\wedge}{\mathit{e}}}:\right\rangle^2}{\left\langle :\overset{\wedge}{\mathit{e}}^{\dagger}{\overset{\wedge}{\mathit{e}}}:\right\rangle}
\end{align}

\subsection{Cosmological Mandel's $\mathbb{Q}$ Parameter for Squeezed Number State (SNS)}

Now to identify the classical or non-classical nature of Squeezed Number State (SNS) in FRW Universe, in this section, we will compute Mandel’s $\mathbb{Q}$ Parameter for SNS using Eq. (\ref{5.6}). Using Eqs. (\ref{3.4}-\ref{3.5}, \ref{3.11}-\ref{3.17}) the values of $\left\langle :\overset{\wedge}{\mathit{e}}^{\dagger2}{\overset{\wedge}{\mathit{e}}}^2:\right\rangle _{\text{SNS}}$ 

\begin{align} \label{5.1.1}
\left\langle :\overset{\wedge}{\mathit{e}}^{\dagger2}{\overset{\wedge}{\mathit{e}}}^2:\right\rangle _{\text{SNS}}=&\biggr\langle n\biggr|\overset{\wedge}W^{\dagger}\biggr[6\overset{\wedge
}{\mathit{e}} ^{\dagger2 }\overset{\wedge}{\mathit{e}}^2+12\overset{\wedge }{\mathit{e}} ^{\dagger}\overset{\wedge }{\mathit{e}}+3-6\overset{\wedge }{\mathit{e}} ^{\dagger2 }+4\overset{\wedge }{\mathit{e}} ^{\dagger3}\overset{\wedge }{\mathit{e}}-6\overset{\wedge }{\mathit{e}} ^2+4\overset{\wedge }{\mathit{e}} ^{\dagger }\overset{\wedge }{\mathit{e}}^3\nonumber\\
&+\overset{\wedge }{\mathit{e}} ^{\dagger 4}+\overset{\wedge }{\mathit{e}}^4\biggr]\biggr[\mathcal{G}^{12}(\mathbf{t})\Phi^2\left(\mathbf{t}\right)\Phi^{*2}\left(\mathbf{t}\right)\dot{\Phi }^2\left(\mathbf{t}_0\right)\dot{\Phi }^{*2}\left(\mathbf{t}_0\right)\nonumber\\
&+\mathcal{G}^{12}(\mathbf{t})\dot{\Phi}^2\left(\mathbf{t}\right)\dot{\Phi}^{*2}\left(\mathbf{t}\right){\Phi }^2\left(\mathbf{t}_0\right){\Phi }^{*2}\left(\mathbf{t}_0\right)\nonumber\\
&-\mathcal{G}^{12}(\mathbf{t})\biggr({\Phi}{^2}\left(\mathbf{t}\right){\Phi}{^*}\left(\mathbf{t}\right)\dot{\Phi}{}\left(\mathbf{t}\right){\Phi}{}\left(\mathbf{t}_0\right)\dot{\Phi}{^*}\left(\mathbf{t}_0\right)\dot{\Phi}{^2}\left(\mathbf{t}_0\right) \nonumber\\
&+ {\Phi}{^2}\left(\mathbf{t}\right){\Phi}{^*}\left(\mathbf{t}\right)\dot{\Phi}{}\left(\mathbf{t}\right){\Phi}{}\left(\mathbf{t}_0\right)\dot{\Phi}^{*2}\left(\mathbf{t}_0\right)\dot{\Phi}{}\left(\mathbf{t}_0\right) \nonumber\\
&+
{\Phi}{}\left(\mathbf{t}\right){\Phi}^{*2}\left(\mathbf{t}\right)\dot{\Phi}{}\left(\mathbf{t}\right){\Phi}{}\left(\mathbf{t}_0\right)\dot{\Phi}{^*}\left(\mathbf{t}_0\right)\dot{\Phi}{^2}\left(\mathbf{t}_0\right) \nonumber\\
&+ {\Phi}{}\left(\mathbf{t}\right){\Phi}^{*2}\left(\mathbf{t}\right)\dot{\Phi}{}\left(\mathbf{t}\right){\Phi}{}\left(\mathbf{t}_0\right)\dot{\Phi}^{*2}\left(\mathbf{t}_0\right)\dot{\Phi}{}\left(\mathbf{t}_0\right)\biggr) \nonumber\\
&-\mathcal{G}^{12}(\mathbf{t})\biggr({\Phi}^{}\left(\mathbf{t}\right)\dot{\Phi}^{*}\left(\mathbf{t}\right)\dot{\Phi}^{2}\left(\mathbf{t}\right){\Phi }^{2}\left(\mathbf{t}_0\right){\Phi }^{*}\left(\mathbf{t}_0\right)\dot{\Phi}^{}\left(\mathbf{t}_0\right) \nonumber\\
& + {\Phi}^{}\left(\mathbf{t}\right)\dot{\Phi}^{*}\left(\mathbf{t}\right)\dot{\Phi}^{2}\left(\mathbf{t}\right){\Phi }^{}\left(\mathbf{t}_0\right){\Phi }^{*2}\left(\mathbf{t}_0\right)\dot{\Phi}^{}\left(\mathbf{t}_0\right) \nonumber\\
& +{\Phi}^{}\left(\mathbf{t}\right)\dot{\Phi}^{*2}\left(\mathbf{t}\right)\dot{\Phi}^{}\left(\mathbf{t}\right){\Phi }^{2}\left(\mathbf{t}_0\right){\Phi }^{*}\left(\mathbf{t}_0\right)\dot{\Phi}^{}\left(\mathbf{t}_0\right) \nonumber\\
& + {\Phi}^{}\left(\mathbf{t}\right)\dot{\Phi}^{*2}\left(\mathbf{t}\right)\dot{\Phi}^{}\left(\mathbf{t}\right){\Phi }^{}\left(\mathbf{t}_0\right){\Phi }^{*2}\left(\mathbf{t}_0\right)\dot{\Phi}^{}\left(\mathbf{t}_0\right)\biggr)\nonumber\\
&+\mathcal{G}^{12}(\mathbf{t})\biggr({\Phi}^{2}\left(\mathbf{t}\right)\dot{\Phi}^{*2}\left(\mathbf{t}\right)\dot{\Phi}^{2}\left(\mathbf{t}_0\right){\Phi }^{*2}\left(\mathbf{t}_0\right) \nonumber\\
& + \dot{\Phi}^{2}\left(\mathbf{t}\right){\Phi }^{*2}\left(\mathbf{t}\right){\Phi}^{2}\left(\mathbf{t}_0\right)\dot{\Phi}^{*2}\left(\mathbf{t}_0\right) \nonumber\\
& +{\Phi}^{}\left(\mathbf{t}\right)\dot{\Phi}^{}\left(\mathbf{t}\right){\Phi}^{*}\left(\mathbf{t}\right)\dot{\Phi}^{*}\left(\mathbf{t}\right){\Phi}^{}\left(\mathbf{t}_0\right)\dot{\Phi}^{}\left(\mathbf{t}_0\right){\Phi}^{*}\left(\mathbf{t}_0\right)\dot{\Phi}^{*}\left(\mathbf{t}_0\right)\nonumber\\
& +{\Phi}^{}\left(\mathbf{t}\right)\dot{\Phi}^{}\left(\mathbf{t}\right){\Phi}^{*}\left(\mathbf{t}\right)\dot{\Phi}^{*}\left(\mathbf{t}\right)\dot{\Phi}^{}\left(\mathbf{t}_0\right){\Phi}^{}\left(\mathbf{t}_0\right)\dot{\Phi}^{*}\left(\mathbf{t}_0\right){\Phi}^{*}\left(\mathbf{t}_0\right)\nonumber\\
& +{\Phi}^{}\left(\mathbf{t}\right)\dot{\Phi}^{}\left(\mathbf{t}\right){\Phi}^{*}\left(\mathbf{t}\right)\dot{\Phi}^{*}\left(\mathbf{t}\right){\Phi}^{2}\left(\mathbf{t}_0\right){\Phi}^{*2}\left(\mathbf{t}_0\right)\nonumber\\
&+{\Phi}^{}\left(\mathbf{t}\right)\dot{\Phi}^{}\left(\mathbf{t}\right){\Phi}^{*}\left(\mathbf{t}\right)\dot{\Phi}^{*}\left(\mathbf{t}\right)\dot{\Phi}^{2}\left(\mathbf{t}_0\right)\dot{\Phi}^{*2}\biggr)\biggr]\overset{\wedge}W\biggr|n\biggr\rangle,
\end{align}

in Eq. (\ref{5.1.1}) by considering 

\begin{align} \label{5.1.2}
\mathbb{X}=&{\Phi}^{}\left(\mathbf{t}\right)\dot{\Phi}^{*}\left(\mathbf{t}\right)\dot{\Phi}^{2}\left(\mathbf{t}\right){\Phi }^{2}\left(\mathbf{t}_0\right){\Phi }^{*}\left(\mathbf{t}_0\right)\dot{\Phi}^{}\left(\mathbf{t}_0\right) + {\Phi}^{}\left(\mathbf{t}\right)\dot{\Phi}^{*}\left(\mathbf{t}\right)\dot{\Phi}^{2}\left(\mathbf{t}\right){\Phi }^{}\left(\mathbf{t}_0\right){\Phi }^{*2}\left(\mathbf{t}_0\right)\dot{\Phi}^{}\left(\mathbf{t}_0\right) \nonumber\\
& +{\Phi}^{}\left(\mathbf{t}\right)\dot{\Phi}^{*2}\left(\mathbf{t}\right)\dot{\Phi}^{}\left(\mathbf{t}\right){\Phi }^{2}\left(\mathbf{t}_0\right){\Phi }^{*}\left(\mathbf{t}_0\right)\dot{\Phi}^{}\left(\mathbf{t}_0\right) + {\Phi}^{}\left(\mathbf{t}\right)\dot{\Phi}^{*2}\left(\mathbf{t}\right)\dot{\Phi}^{}\left(\mathbf{t}\right){\Phi }^{}\left(\mathbf{t}_0\right){\Phi }^{*2}\left(\mathbf{t}_0\right)\dot{\Phi}^{}\left(\mathbf{t}_0\right),
\end{align}

\begin{align} \label{5.1.3}
\mathbb{Y}=&{\Phi}{^2}\left(\mathbf{t}\right){\Phi}{^*}\left(\mathbf{t}\right)\dot{\Phi}{}\left(\mathbf{t}\right){\Phi}{}\left(\mathbf{t}_0\right)\dot{\Phi}{^*}\left(\mathbf{t}_0\right)\dot{\Phi}{^2}\left(\mathbf{t}_0\right) + {\Phi}{^2}\left(\mathbf{t}\right){\Phi}{^*}\left(\mathbf{t}\right)\dot{\Phi}{}\left(\mathbf{t}\right){\Phi}{}\left(\mathbf{t}_0\right)\dot{\Phi}^{*2}\left(\mathbf{t}_0\right)\dot{\Phi}{}\left(\mathbf{t}_0\right) \nonumber\\
&+
{\Phi}{}\left(\mathbf{t}\right){\Phi}^{*2}\left(\mathbf{t}\right)\dot{\Phi}{}\left(\mathbf{t}\right){\Phi}{}\left(\mathbf{t}_0\right)\dot{\Phi}{^*}\left(\mathbf{t}_0\right)\dot{\Phi}{^2}\left(\mathbf{t}_0\right) + {\Phi}{}\left(\mathbf{t}\right){\Phi}^{*2}\left(\mathbf{t}\right)\dot{\Phi}{}\left(\mathbf{t}\right){\Phi}{}\left(\mathbf{t}_0\right)\dot{\Phi}^{*2}\left(\mathbf{t}_0\right)\dot{\Phi}{}\left(\mathbf{t}_0\right), 
\end{align}

\begin{align} \label{5.1.4}
\mathbb{Z}=&{\Phi}^{2}\left(\mathbf{t}\right)\dot{\Phi}^{*2}\left(\mathbf{t}\right)\dot{\Phi}^{2}\left(\mathbf{t}_0\right){\Phi }^{*2}\left(\mathbf{t}_0\right) + \dot{\Phi}^{2}\left(\mathbf{t}\right){\Phi }^{*2}\left(\mathbf{t}\right){\Phi}^{2}\left(\mathbf{t}_0\right)\dot{\Phi}^{*2}\left(\mathbf{t}_0\right) \nonumber\\
& +{\Phi}^{}\left(\mathbf{t}\right)\dot{\Phi}^{}\left(\mathbf{t}\right){\Phi}^{*}\left(\mathbf{t}\right)\dot{\Phi}^{*}\left(\mathbf{t}\right){\Phi}^{}\left(\mathbf{t}_0\right)\dot{\Phi}^{}\left(\mathbf{t}_0\right){\Phi}^{*}\left(\mathbf{t}_0\right)\dot{\Phi}^{*}\left(\mathbf{t}_0\right)\nonumber\\
& +{\Phi}^{}\left(\mathbf{t}\right)\dot{\Phi}^{}\left(\mathbf{t}\right){\Phi}^{*}\left(\mathbf{t}\right)\dot{\Phi}^{*}\left(\mathbf{t}\right)\dot{\Phi}^{}\left(\mathbf{t}_0\right){\Phi}^{}\left(\mathbf{t}_0\right)\dot{\Phi}^{*}\left(\mathbf{t}_0\right){\Phi}^{*}\left(\mathbf{t}_0\right)\nonumber\\
& +{\Phi}^{}\left(\mathbf{t}\right)\dot{\Phi}^{}\left(\mathbf{t}\right){\Phi}^{*}\left(\mathbf{t}\right)\dot{\Phi}^{*}\left(\mathbf{t}\right){\Phi}^{2}\left(\mathbf{t}_0\right){\Phi}^{*2}\left(\mathbf{t}_0\right)+{\Phi}^{}\left(\mathbf{t}\right)\dot{\Phi}^{}\left(\mathbf{t}\right){\Phi}^{*}\left(\mathbf{t}\right)\dot{\Phi}^{*}\left(\mathbf{t}\right)\dot{\Phi}^{2}\left(\mathbf{t}_0\right)\dot{\Phi}^{*2}
\end{align}

using Eqs. (\ref{5.1.2}-\ref{5.1.4}) in Eqs. (\ref{5.1.1}) as

\begin{align} \label{5.1.5}
\left\langle :\overset{\wedge}{\mathit{e}}^{\dagger2}{\overset{\wedge}{\mathit{e}}}^2:\right\rangle _{\text{SNS}}=&\biggr\langle n\biggr|\overset{\wedge}W^{\dagger}\biggr[6\overset{\wedge
}{\mathit{e}} ^{\dagger2 }\overset{\wedge}{\mathit{e}}^2+12\overset{\wedge }{\mathit{e}} ^{\dagger}\overset{\wedge }{\mathit{e}}+3-6\overset{\wedge }{\mathit{e}} ^{\dagger2 }+4\overset{\wedge }{\mathit{e}} ^{\dagger3}\overset{\wedge }{\mathit{e}}-6\overset{\wedge }{\mathit{e}} ^2+4\overset{\wedge }{\mathit{e}} ^{\dagger }\overset{\wedge }{\mathit{e}}^3\nonumber\\
&+\overset{\wedge }{\mathit{e}} ^{\dagger 4}+\overset{\wedge }{\mathit{e}}^4\biggr]\biggr[\mathcal{G}^{12}(\mathbf{t})\Phi^2\left(\mathbf{t}\right)\Phi^{*2}\left(\mathbf{t}\right)\dot{\Phi }^2\left(\mathbf{t}_0\right)\dot{\Phi }^{*2}\left(\mathbf{t}_0\right)\nonumber\\
&+\mathcal{G}^{12}(\mathbf{t})\dot{\Phi}^2\left(\mathbf{t}\right)\dot{\Phi}^{*2}\left(\mathbf{t}\right){\Phi }^2\left(\mathbf{t}_0\right){\Phi }^{*2}\left(\mathbf{t}_0\right)\nonumber\\
&-\mathbb{X}\mathcal{G}^{12}(\mathbf{t})-\mathbb{Y}\mathcal{G}^{12}(\mathbf{t})+\mathbb{Z}\mathcal{G}^{12}(\mathbf{t})\biggr]\overset{\wedge}W\biggr|n\biggr\rangle,
\end{align}

the value of $\left\langle :\overset{\wedge}{\mathit{e}}^{\dagger2}{\overset{\wedge}{\mathit{e}}}^2:\right\rangle _{\text{SNS}}$ can be further computed using Eqs. (\ref{3.11}-\ref{3.17}, \ref{4.7}-\ref{4.10}) in Eq. (\ref{5.1.5})

\begin{align} \label{5.1.6}
\left\langle :\overset{\wedge}{\mathit{e}}^{\dagger2}{\overset{\wedge}{\mathit{e}}}^2:\right\rangle _{\text{SNS}}=&\left(\frac{1}{16m^{4}\mathbf{t}^4\mathbf{t}_0^4}+\frac{1}{16m^{4}}+\frac{3}{8m^{4}\mathbf{t}^2\mathbf{t}_0^2}-\frac{1}{4m^{4}\mathbf{t}^3\mathbf{t}_0^3}-\frac{1}{4m^{4}\mathbf{t}\mathbf{t}_0}\right)\biggr[3\nonumber\\
&+\text{Cosh}^4\rho
(6n^2-6n)+\text{Sinh}^4\rho \left(6n^2+18n+12\right)\nonumber\\
&+\text{Cosh}^2\rho \text{Sinh}^2\rho \left(36n^2+36n+12\right)+\text{Cosh}^3\rho \text{Sinh$\rho
$}\left(24n^2\right)\nonumber\\
&+\text{Cosh$\rho $} \text{Sinh}^3\rho \left(24n^2+48n+24\right)+\text{Cosh$\rho $} \text{Sinh$\rho $}(24n+12)\nonumber\\
&+\text{Sinh}^2\rho (12n+12)+\text{Cosh}^2\rho(12n)\biggr].
\end{align}

using Eqs. (\ref{3.11}-\ref{3.17}, \ref{4.7}-\ref{4.10}) the values of $\left\langle :\overset{\wedge}{\mathit{e}}^{\dagger}{\overset{\wedge}{\mathit{e}}}:\right\rangle _{\text{SNS}}$ 

\begin{align} \label{5.1.7}
\left\langle :\overset{\wedge}{\mathit{e}}^{\dagger}{\overset{\wedge}{\mathit{e}}}:\right\rangle _{\text{SNS}}=&\biggr\langle n\biggr|\overset{\wedge}W^{\dagger}\biggr[2\overset{\wedge }{\mathit{e}} ^{\dagger }\overset{\wedge }{\mathit{e}}+1-\overset{\wedge }{\mathit{e}} ^{\dagger2 }-\overset{\wedge }{\mathit{e}} ^2\biggr]\biggr[\mathcal{G}^{6}(\mathbf{t}){\Phi}\left(\mathbf{t}\right){\Phi}^*\left(\mathbf{t}\right)\dot{\Phi }\left(\mathbf{t}_0\right)\dot{\Phi}\left(\mathbf{t}_0\right)\nonumber\\
&+\mathcal{G}^{6}(\mathbf{t})\dot{\Phi}\left(\mathbf{t}\right)\dot{\Phi}^*\left(\mathbf{t}\right){\Phi }\left(\mathbf{t}_0\right){\Phi}\left(\mathbf{t}_0\right)\nonumber\\
&-\mathcal{G}^{6}(\mathbf{t})\dot{\Phi}\left(\mathbf{t}\right){\Phi}^*\left(\mathbf{t}\right){\Phi }\left(\mathbf{t}_0\right)\dot{\Phi}^*\left(\mathbf{t}_0\right)\nonumber\\
&-\mathcal{G}^{6}(\mathbf{t}){\Phi}\left(\mathbf{t}\right)\dot{\Phi}^*\left(\mathbf{t}\right){\Phi }^*\left(\mathbf{t}_0\right)\dot{\Phi}\left(\mathbf{t}_0\right)\biggr]\overset{\wedge}W\biggr|n\biggr\rangle,
\end{align}

the value of $\left\langle :\overset{\wedge}{\mathit{e}}^{\dagger}{\overset{\wedge}{\mathit{e}}}:\right\rangle _{\text{SNS}}$ can be further computed using Using Eqs. (\ref{3.11}-\ref{3.17}, \ref{4.7}-\ref{4.10}) in Eq. (\ref{5.1.7})

\begin{align} \label{5.1.8}
\left\langle :\overset{\wedge}{\mathit{e}}^{\dagger}{\overset{\wedge}{\mathit{e}}}:\right\rangle _{\text{SNS}}=&\left(\frac{1}{4m^{2}\mathbf{t}^2\mathbf{t}_0^2}+\frac{1}{4m^{2}}-\frac{1}{2m^{2}\mathbf{t}\mathbf{t}_0}\right)[\text{Cosh}^2\rho (2n)+\text{Sinh}^2\rho
(2n+2)\nonumber\\
&+\text{Cosh$\rho $} \text{Sinh$\rho $}(4n+2)+1].
\end{align}

Using Eqs. (\ref{3.11}-\ref{3.17}, \ref{4.7}-\ref{4.10}) the values of $\left\langle :\overset{\wedge}{\mathit{e}}^{\dagger}{\overset{\wedge}{\mathit{e}}}:\right\rangle^2_{\text{SNS}}$ 

\begin{align} \label{5.1.9}
\left\langle :\overset{\wedge}{\mathit{e}}^{\dagger}{\overset{\wedge}{\mathit{e}}}:\right\rangle^2_{\text{SNS}}=&\biggr\langle n\biggr|\overset{\wedge}W^{\dagger}\biggr[2\overset{\wedge }{\mathit{e}} ^{\dagger }\overset{\wedge }{\mathit{e}}+1-\overset{\wedge }{\mathit{e}} ^{\dagger2 }-\overset{\wedge }{\mathit{e}} ^2\biggr]^2\biggr[\mathcal{G}^{6}(\mathbf{t}){\Phi}\left(\mathbf{t}\right){\Phi}^*\left(\mathbf{t}\right)\dot{\Phi }\left(\mathbf{t}_0\right)\dot{\Phi}\left(\mathbf{t}_0\right)\nonumber\\
&+\mathcal{G}^{6}(\mathbf{t})\dot{\Phi}\left(\mathbf{t}\right)\dot{\Phi}^*\left(\mathbf{t}\right){\Phi }\left(\mathbf{t}_0\right){\Phi}\left(\mathbf{t}_0\right)\nonumber\\
&-\mathcal{G}^{6}(\mathbf{t})\dot{\Phi}\left(\mathbf{t}\right){\Phi}^*\left(\mathbf{t}\right){\Phi }\left(\mathbf{t}_0\right)\dot{\Phi}^*\left(\mathbf{t}_0\right)\nonumber\\
&-\mathcal{G}^{6}(\mathbf{t}){\Phi}\left(\mathbf{t}\right)\dot{\Phi}^*\left(\mathbf{t}\right){\Phi }^*\left(\mathbf{t}_0\right)\dot{\Phi}\left(\mathbf{t}_0\right)\biggr]^2\overset{\wedge}W\biggr|n\biggr\rangle,
\end{align}

the value of $\left\langle :\overset{\wedge}{\mathit{e}}^{\dagger}{\overset{\wedge}{\mathit{e}}}:\right\rangle^2_{\text{SNS}}$ can be further computed using Using Eqs. (\ref{3.11}-\ref{3.17}, \ref{4.7}-\ref{4.10}) in Eq. (\ref{5.1.9})

\begin{align} \label{5.1.10}
\left\langle :\overset{\wedge}{\mathit{e}}^{\dagger}{\overset{\wedge}{\mathit{e}}}:\right\rangle^2_{\text{SNS}}=&\left(\frac{1}{16m^{4}\mathbf{t}^4\mathbf{t}_0^4}+\frac{1}{16m^{4}}+\frac{3}{8m^{4}\mathbf{t}^2\mathbf{t}_0^2}-\frac{1}{4m^{4}\mathbf{t}^3\mathbf{t}_0^3}-\frac{1}{4m^{4}\mathbf{t}\mathbf{t}_0}\right)\biggr[1\nonumber\\
&+\text{Cosh}^4\rho
\left(4n^2\right)+\text{Sinh}^4\rho \left(4n^2+8n+4\right)\nonumber\\
&+\text{Cosh}^2\rho\text{Sinh}^2\rho \left(24n^2+24n+4\right)+\text{Cosh}^3\rho \text{Sinh$\rho
$}(16n^2\nonumber\\
&+8n)+\text{Cosh$\rho $} \text{Sinh}^3\rho \left(16n^2+24n+8\right)\nonumber\\
&+\text{Cosh$\rho $} \text{Sinh$\rho $}(8n+4)+\text{Sinh}^2\rho (4n+4)+\text{Cosh}^2\rho
(4n)\biggr].
\end{align}

Substituting the values of $\left\langle :\overset{\wedge}{\mathit{e}}^{\dagger2}{\overset{\wedge}{\mathit{e}}}^2:\right\rangle _{\text{SNS}}$, $\left\langle :\overset{\wedge}{\mathit{e}}^{\dagger}{\overset{\wedge}{\mathit{e}}}:\right\rangle _{\text{SNS}}$ and $\left\langle :\overset{\wedge}{\mathit{e}}^{\dagger}{\overset{\wedge}{\mathit{e}}}:\right\rangle^2_{\text{SNS}}$ in Eq. (\ref{5.6}), Cosmological Mandel's $\mathbb{Q}$ Parameter for squeezed number state is

\begin{align} \label{5.1.11}
\mathbb{Q}_{\text{SNS}}=&\left(\frac{1}{4m^{2}\mathbf{t}^2\mathbf{t}_0^2}+\frac{1}{4m^{2}}-\frac{1}{2m^{2}\mathbf{t}\mathbf{t}_0}\right)\biggr[2+\text{Cosh}^4\rho (2n^2-6n)\nonumber\\
&+\text{Sinh}^4\rho \left(2n^2+10n+8\right)+\text{Cosh}^2\rho
 \text{Sinh}^2\rho \left(12n^2+12n+8\right)\nonumber\\
&+\text{Cosh}^3\rho\text{Sinh$\rho $}\left(8n^2-8n\right)+\text{Cosh$\rho $}\text{Sinh}^3\rho \left(8n^2+24n+16\right)\nonumber\\
&+\text{Cosh$\rho
$}\text{Sinh$\rho $}(16n+8)+\text{Sinh}^2\rho (8n+8)+\text{Cosh}^2\rho
(8n)\biggr]{/}\biggr[\text{Cosh}^2\rho (2n)\nonumber\\
&+\text{Sinh}^2\rho
(2n+2)+\text{Cosh$\rho $} \text{Sinh$\rho $}(4n+2)+1\biggr]
\end{align}

\begin{table}[h]\label{table 1}
\caption{Numerical values of Mandel's $\mathbf{Q}$ Parameter while n=1, $\mathbf{t}_0=1$ for various values of $\rho$ and $\triangle\mathbf{t}$}
\label{tab:table_1}
\begin{tabular}{@{}llllllll@{}}
\toprule
$\rho$ & $\triangle\mathbf{t}=0.1 $  & $\triangle\mathbf{t}=0.2 $ & $\triangle\mathbf{t}=0.3 $ & $ \triangle\mathbf{t}=0.4 $ & $ \triangle\mathbf{t}=0.5$ & $\triangle\mathbf{t}=1 $ & $\triangle\mathbf{t}=5 $\\
\midrule
0.002 & 0.0041 & 0.0139 & 0.0267 & 0.0410 & 0.0558 & 0.1255 & 0.3213 \\ 
0.004 & 0.0042 & 0.0140 & 0.0268 & 0.0411 & 0.0560 & 0.1260 & 0.3226 \\ 
0.006 & 0.0042 & 0.0141 & 0.0269 & 0.0413 & 0.0562 & 0.1265 & 0.3239 \\ 
0.008 & 0.0042 & 0.0141 & 0.0271 & 0.0415 & 0.0565 & 0.1270 & 0.3252 \\ 
0.010 & 0.0042 & 0.0142 & 0.0272 & 0.0416 & 0.0567 & 0.1275 & 0.3265 \\ 
0.020 & 0.0043 & 0.0145 & 0.0277 & 0.0425 & 0.0578 & 0.1301 & 0.3331 \\ 
0.040 & 0.0045 & 0.0150 & 0.0288 & 0.0442 & 0.0602 & 0.1354 & 0.3467 \\ 
0.060 & 0.0047 & 0.0157 & 0.0300 & 0.0460 & 0.0626 & 0.1409 & 0.3608 \\ 
0.080 & 0.0048 & 0.0163 & 0.0312 & 0.0479 & 0.0652 & 0.1467 & 0.3755 \\ 
0.100 & 0.0050 & 0.0170 & 0.0325 & 0.0499 & 0.0679 & 0.1527 & 0.3908 \\ 
0.200 & 0.0062 & 0.0207 & 0.0397 & 0.0609 & 0.0829 & 0.1865 & 0.4774 \\ 
0.400 & 0.0092 & 0.0309 & 0.0593 & 0.0908 & 0.1236 & 0.2782 & 0.7122 \\ 
0.600 & 0.0137 & 0.0461 & 0.0884 & 0.1355 & 0.1845 & 0.4150 & 1.0624 \\ 
0.800 & 0.0205 & 0.0688 & 0.1319 & 0.2022 & 0.2752 & 0.6191 & 1.5850 \\ 
1.000 & 0.0305 & 0.1026 & 0.1967 & 0.3016 & 0.4105 & 0.9236 & 2.3645 \\ 
1.200 & 0.0456 & 0.1531 & 0.2935 & 0.4499 & 0.6124 & 1.3779 & 3.5274 \\ 
1.400 & 0.0680 & 0.2284 & 0.4379 & 0.6712 & 0.9136 & 2.0556 & 5.2623 \\ 
1.600 & 0.1014 & 0.3407 & 0.6532 & 1.0013 & 1.3629 & 3.0666 & 7.8504 \\ 
1.800 & 0.1512 & 0.5083 & 0.9745 & 1.4938 & 2.0332 & 4.5748 & 11.7114 \\ 
2.000 & 0.2256 & 0.7583 & 1.4538 & 2.2285 & 3.0332 & 6.8248 & 17.4714 \\ 
\botrule
\end{tabular}
\end{table}

\begin{table}[h]\label{table 2}
\caption{Numerical values of Mandel's $\mathbf{Q}$ Parameter while n=2, $\mathbf{t}_0=1$ for various values of $\rho$ and $\triangle\mathbf{t}$}
\label{tab:table_2}
\begin{tabular}{@{}llllllll@{}}
\toprule
$\rho$ & $\triangle\mathbf{t}=0.1 $  & $\triangle\mathbf{t}=0.2 $ & $\triangle\mathbf{t}=0.3 $ & $ \triangle\mathbf{t}=0.4 $ & $ \triangle\mathbf{t}=0.5$ & $\triangle\mathbf{t}=1 $ & $\triangle\mathbf{t}=5 $\\
\midrule
0.002 & 0.0058 & 0.0195 & 0.0374 & 0.0574 & 0.0781 & 0.1757 & 0.4498 \\ 
0.004 & 0.0058 & 0.0196 & 0.0376 & 0.0576 & 0.0784 & 0.1764 & 0.4516 \\ 
0.006 & 0.0059 & 0.0197 & 0.0377 & 0.0578 & 0.0787 & 0.1771 & 0.4534 \\ 
0.008 & 0.0059 & 0.0198 & 0.0379 & 0.0581 & 0.0790 & 0.1778 & 0.4552 \\ 
0.010 & 0.0059 & 0.0198 & 0.0380 & 0.0583 & 0.0793 & 0.1785 & 0.4570 \\ 
0.020 & 0.0060 & 0.0202 & 0.0388 & 0.0595 & 0.0810 & 0.1821 & 0.4663 \\ 
0.040 & 0.0063 & 0.0211 & 0.0404 & 0.0619 & 0.0843 & 0.1896 & 0.4853 \\ 
0.060 & 0.0065 & 0.0219 & 0.0420 & 0.0644 & 0.0877 & 0.1973 & 0.5051 \\ 
0.080 & 0.0068 & 0.0228 & 0.0437 & 0.0671 & 0.0913 & 0.2054 & 0.5257 \\ 
0.100 & 0.0071 & 0.0237 & 0.0455 & 0.0698 & 0.0950 & 0.2137 & 0.5472 \\ 
0.200 & 0.0086 & 0.0290 & 0.0556 & 0.0852 & 0.1160 & 0.2611 & 0.6683 \\ 
0.400 & 0.0129 & 0.0433 & 0.0830 & 0.1272 & 0.1731 & 0.3895 & 0.9970 \\ 
0.600 & 0.0192 & 0.0646 & 0.1238 & 0.1897 & 0.2582 & 0.5810 & 1.4874 \\ 
0.800 & 0.0287 & 0.0963 & 0.1846 & 0.2830 & 0.3852 & 0.8668 & 2.2190 \\ 
1.000 & 0.0427 & 0.1437 & 0.2755 & 0.4222 & 0.5747 & 1.2931 & 3.3103 \\ 
1.200 & 0.0638 & 0.2143 & 0.4109 & 0.6299 & 0.8574 & 1.9291 & 4.9384 \\ 
1.400 & 0.0951 & 0.3198 & 0.6130 & 0.9397 & 1.2790 & 2.8778 & 7.3672 \\ 
1.600 & 0.1419 & 0.4770 & 0.9145 & 1.4019 & 1.9081 & 4.2932 & 10.9906 \\ 
1.800 & 0.2117 & 0.7116 & 1.3643 & 2.0913 & 2.8465 & 6.4047 & 16.3960 \\ 
2.000 & 0.3159 & 1.0616 & 2.0353 & 3.1199 & 4.2465 & 9.5547 & 24.4600 \\ 
\botrule
\end{tabular}
\end{table}

\begin{table}[h]\label{table 3}
\caption{Numerical values of Mandel's $\mathbf{Q}$ Parameter while n=3, $\mathbf{t}_0=1$ for various values of $\rho$ and $\triangle\mathbf{t}$}
\label{tab:table_3}
\begin{tabular}{@{}llllllll@{}}
\toprule
$\rho$ & $\triangle\mathbf{t}=0.1 $  & $\triangle\mathbf{t}=0.2 $ & $\triangle\mathbf{t}=0.3 $ & $ \triangle\mathbf{t}=0.4 $ & $ \triangle\mathbf{t}=0.5$ & $\triangle\mathbf{t}=1 $ & $\triangle\mathbf{t}=5 $\\
\midrule
0.002 & 0.0077 & 0.0259 & 0.0496 & 0.0761 & 0.1036 & 0.2331 & 0.5967 \\ 
0.004 & 0.0077 & 0.0260 & 0.0498 & 0.0764 & 0.1040 & 0.2340 & 0.5991 \\ 
0.006 & 0.0078 & 0.0261 & 0.0500 & 0.0767 & 0.1044 & 0.2349 & 0.6015 \\ 
0.008 & 0.0078 & 0.0262 & 0.0502 & 0.0770 & 0.1048 & 0.2359 & 0.6039 \\ 
0.010 & 0.0078 & 0.0263 & 0.0504 & 0.0773 & 0.1053 & 0.2368 & 0.6063 \\ 
0.020 & 0.0080 & 0.0268 & 0.0515 & 0.0789 & 0.1074 & 0.2416 & 0.6185 \\ 
0.040 & 0.0083 & 0.0279 & 0.0536 & 0.0821 & 0.1118 & 0.2515 & 0.6438 \\ 
0.060 & 0.0087 & 0.0291 & 0.0558 & 0.0855 & 0.1163 & 0.2617 & 0.6701 \\ 
0.080 & 0.0090 & 0.0303 & 0.0580 & 0.0890 & 0.1211 & 0.2724 & 0.6974 \\ 
0.100 & 0.0094 & 0.0315 & 0.0604 & 0.0926 & 0.1260 & 0.2835 & 0.7259 \\ 
0.200 & 0.0114 & 0.0385 & 0.0738 & 0.1131 & 0.1539 & 0.3463 & 0.8866 \\ 
0.400 & 0.0171 & 0.0574 & 0.1101 & 0.1687 & 0.2296 & 0.5166 & 1.3226 \\ 
0.600 & 0.0255 & 0.0856 & 0.1642 & 0.2517 & 0.3426 & 0.7707 & 1.9731 \\ 
0.800 & 0.0380 & 0.1278 & 0.2449 & 0.3754 & 0.5110 & 1.1498 & 2.9435 \\ 
1.000 & 0.0567 & 0.1906 & 0.3654 & 0.5601 & 0.7624 & 1.7153 & 4.3912 \\ 
1.200 & 0.0846 & 0.2843 & 0.5451 & 0.8356 & 1.1373 & 2.5590 & 6.5509 \\ 
1.400 & 0.1262 & 0.4242 & 0.8132 & 1.2465 & 1.6967 & 3.8175 & 9.7728 \\ 
1.600 & 0.1883 & 0.6328 & 1.2131 & 1.8596 & 2.5311 & 5.6951 & 14.5793 \\ 
1.800 & 0.2809 & 0.9440 & 1.8098 & 2.7742 & 3.7760 & 8.4960 & 21.7498 \\ 
2.000 & 0.4190 & 1.4083 & 2.6999 & 4.1386 & 5.6331 & 12.6746 & 32.4469 \\ 
\botrule
\end{tabular}
\end{table}

\begin{table}[h]\label{table 4}
\caption{Numerical values of Mandel's $\mathbf{Q}$ Parameter while n=4, $\mathbf{t}_0=1$ for various values of $\rho$ and $\triangle\mathbf{t}$}
\label{tab:table_4}
\begin{tabular}{@{}llllllll@{}}
\toprule
$\rho$ & $\triangle\mathbf{t}=0.1 $  & $\triangle\mathbf{t}=0.2 $ & $\triangle\mathbf{t}=0.3 $ & $ \triangle\mathbf{t}=0.4 $ & $ \triangle\mathbf{t}=0.5$ & $\triangle\mathbf{t}=1 $ & $\triangle\mathbf{t}=5 $\\
\midrule
0.002 & 0.0097 & 0.0325 & 0.0624 & 0.0956 & 0.1301 & 0.2928 & 0.7497 \\ 
0.004 & 0.0097 & 0.0327 & 0.0626 & 0.0960 & 0.1307 & 0.2940 & 0.7527 \\ 
0.006 & 0.0098 & 0.0328 & 0.0629 & 0.0964 & 0.1312 & 0.2952 & 0.7557 \\ 
0.008 & 0.0098 & 0.0329 & 0.0631 & 0.0968 & 0.1317 & 0.2964 & 0.7587 \\ 
0.010 & 0.0098 & 0.0331 & 0.0634 & 0.0972 & 0.1322 & 0.2976 & 0.7618 \\
0.020 & 0.0100 & 0.0337 & 0.0647 & 0.0991 & 0.1349 & 0.3036 & 0.7771 \\ 
0.040 & 0.0104 & 0.0351 & 0.0673 & 0.1032 & 0.1404 & 0.3160 & 0.8089 \\ 
0.060 & 0.0109 & 0.0365 & 0.0701 & 0.1074 & 0.1462 & 0.3289 & 0.8419 \\ 
0.080 & 0.0113 & 0.0380 & 0.0729 & 0.1118 & 0.1521 & 0.3423 & 0.8762 \\ 
0.100 & 0.0118 & 0.0396 & 0.0759 & 0.1163 & 0.1583 & 0.3562 & 0.9120 \\ 
0.200 & 0.0144 & 0.0483 & 0.0927 & 0.1421 & 0.1934 & 0.4351 & 1.1139 \\ 
0.400 & 0.0215 & 0.0721 & 0.1383 & 0.2120 & 0.2885 & 0.6491 & 1.6617 \\ 
0.600 & 0.0320 & 0.1076 & 0.2063 & 0.3162 & 0.4304 & 0.9684 & 2.4790 \\ 
0.800 & 0.0478 & 0.1605 & 0.3077 & 0.4717 & 0.6421 & 1.4446 & 3.6983 \\ 
1.000 & 0.0712 & 0.2395 & 0.4591 & 0.7037 & 0.9578 & 2.1551 & 5.5172 \\ 
1.200 & 0.1063 & 0.3572 & 0.6849 & 1.0498 & 1.4289 & 3.2151 & 8.2306 \\ 
1.400 & 0.1586 & 0.5329 & 1.0217 & 1.5662 & 2.1317 & 4.7964 & 12.2787 \\ 
1.600 & 0.2365 & 0.7950 & 1.5242 & 2.3364 & 3.1801 & 7.1553 & 18.3176 \\ 
1.800 & 0.3529 & 1.1861 & 2.2739 & 3.4856 & 4.7442 & 10.6745 & 27.3267 \\ 
2.000 & 0.5264 & 1.7694 & 3.3922 & 5.1998 & 7.0775 & 15.9244 & 40.7666 \\ 
\botrule
\end{tabular}
\end{table}

\begin{table}[h]\label{table 5}
\caption{Numerical values of Mandel's $\mathbf{Q}$ Parameter while n=5, $\mathbf{t}_0=1$ for various values of $\rho$ and $\triangle\mathbf{t}$}
\label{tab:table_5}
\begin{tabular}{@{}llllllll@{}}
\toprule
$\rho$ & $\triangle\mathbf{t}=0.1 $  & $\triangle\mathbf{t}=0.2 $ & $\triangle\mathbf{t}=0.3 $ & $ \triangle\mathbf{t}=0.4 $ & $ \triangle\mathbf{t}=0.5$ & $\triangle\mathbf{t}=1 $ & $\triangle\mathbf{t}=5 $\\
\midrule
0.002 & 0.0117 & 0.0393 & 0.0753 & 0.1155 & 0.1572 & 0.3537 & 0.9054 \\ 
0.004 & 0.0117 & 0.0395 & 0.0756 & 0.1160 & 0.1578 & 0.3551 & 0.9091 \\ 
0.006 & 0.0118 & 0.0396 & 0.0759 & 0.1164 & 0.1585 & 0.3565 & 0.9127 \\ 
0.008 & 0.0118 & 0.0398 & 0.0763 & 0.1169 & 0.1591 & 0.3580 & 0.9164 \\ 
0.010 & 0.0119 & 0.0399 & 0.0766 & 0.1174 & 0.1597 & 0.3594 & 0.9200 \\ 
0.020 & 0.0121 & 0.0407 & 0.0781 & 0.1197 & 0.1630 & 0.3666 & 0.9386 \\ 
0.040 & 0.0126 & 0.0424 & 0.0813 & 0.1246 & 0.1696 & 0.3816 & 0.9769 \\ 
0.060 & 0.0131 & 0.0441 & 0.0846 & 0.1297 & 0.1765 & 0.3972 & 1.0168 \\ 
0.080 & 0.0137 & 0.0459 & 0.0881 & 0.1350 & 0.1837 & 0.4134 & 1.0583 \\ 
0.100 & 0.0142 & 0.0478 & 0.0917 & 0.1405 & 0.1912 & 0.4303 & 1.1015 \\ 
0.200 & 0.0174 & 0.0584 & 0.1119 & 0.1716 & 0.2336 & 0.5255 & 1.3454 \\ 
0.400 & 0.0259 & 0.0871 & 0.1670 & 0.2560 & 0.3484 & 0.7840 & 2.0070 \\ 
0.600 & 0.0387 & 0.1300 & 0.2491 & 0.3819 & 0.5198 & 1.1696 & 2.9941 \\ 
0.800 & 0.0577 & 0.1939 & 0.3717 & 0.5697 & 0.7755 & 1.7448 & 4.4667 \\ 
1.000 & 0.0860 & 0.2892 & 0.5545 & 0.8499 & 1.1569 & 2.6030 & 6.6636 \\ 
1.200 & 0.1284 & 0.4315 & 0.8272 & 1.2680 & 1.7259 & 3.8832 & 9.9409 \\ 
1.400 & 0.1915 & 0.6437 & 1.2340 & 1.8916 & 2.5747 & 5.7930 & 14.8301 \\ 
1.600 & 0.2857 & 0.9602 & 1.8409 & 2.8219 & 3.8409 & 8.6421 & 22.1238 \\ 
1.800 & 0.4262 & 1.4325 & 2.7463 & 4.2098 & 5.7300 & 12.8926 & 33.0050 \\ 
2.000 & 0.6358 & 2.1370 & 4.0971 & 6.2803 & 8.5482 & 19.2334 & 49.2376 \\  
\botrule
\end{tabular}
\end{table}

\begin{figure}[h]%
\centering
\includegraphics[width=0.9\textwidth]{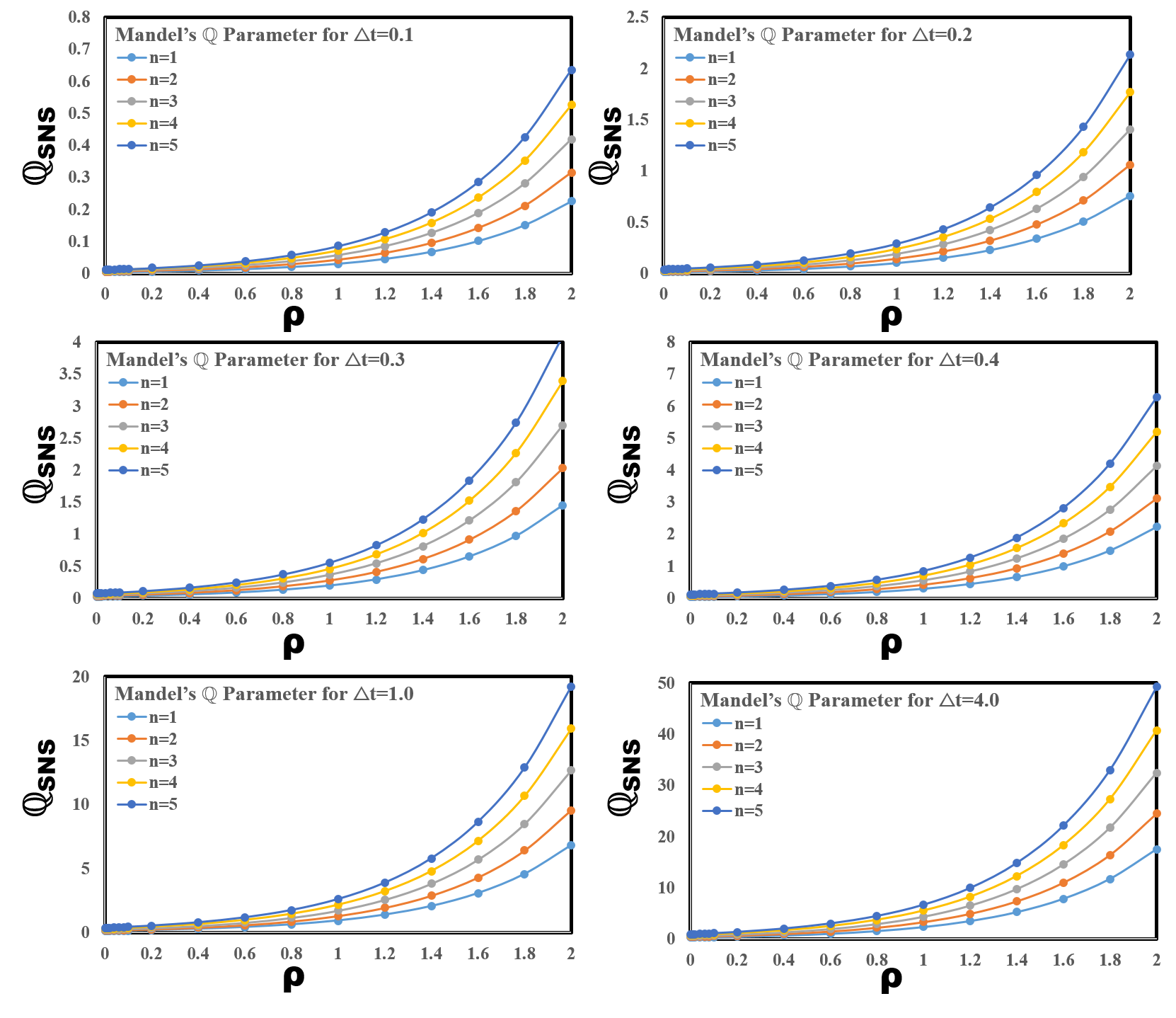}
\caption{Dependence of Mandel's $\mathbf{Q_{\text{SNS}}}$ Parameter on the squeezing parameter ($\rho $) and quantum number (n). The plot illustrates the relationship highlighting the non-classical characteristics as influenced by varying levels of squeezing and quantum states.}
\label{fig:figure_1}
\end{figure}

\begin{figure}[h]%
\centering
\includegraphics[width=0.9\textwidth]{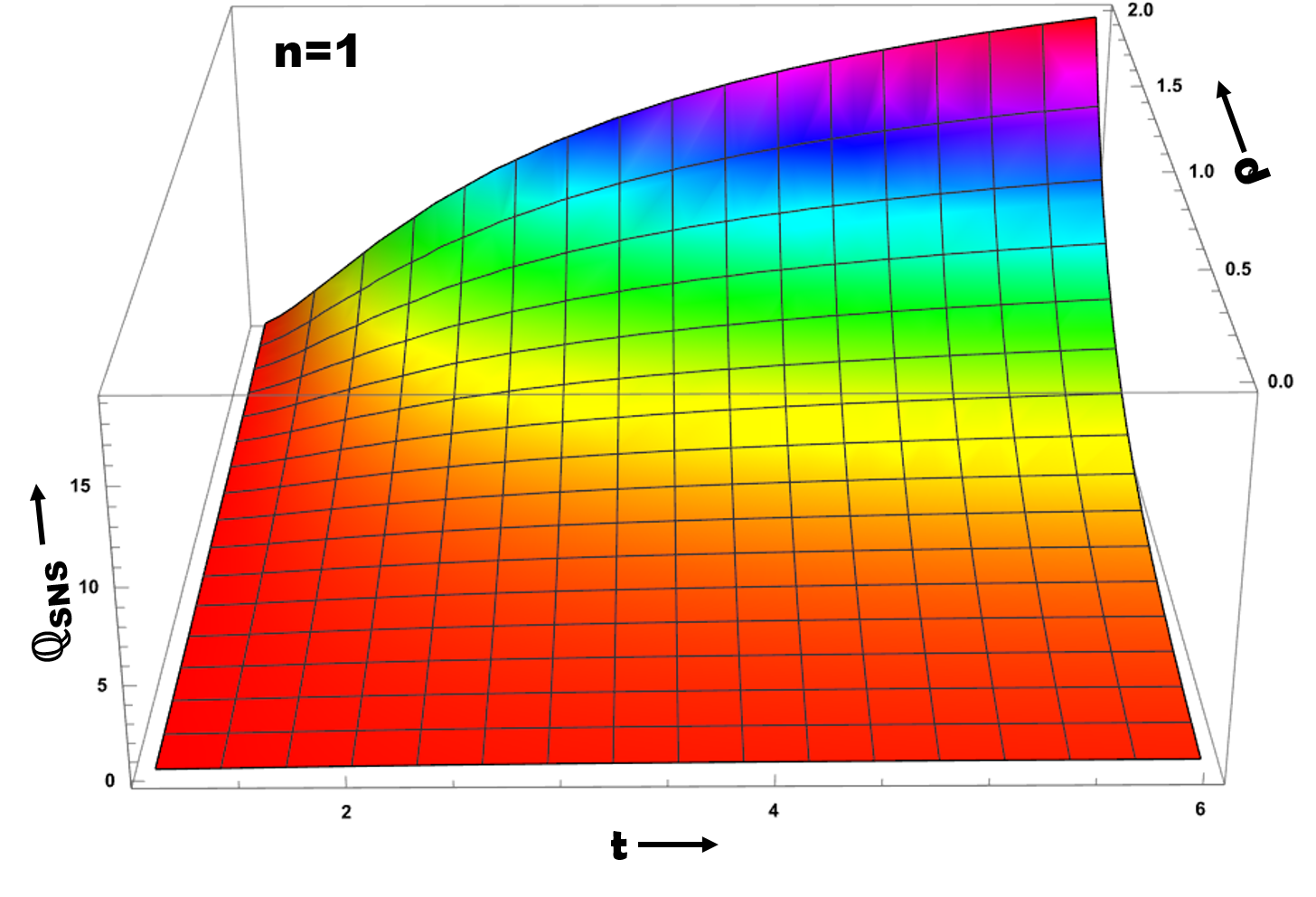}
\caption{3-D representation of the variation in Mandel's $\mathbf{Q_{\text{SNS}}}$ as a function of the squeezing parameter ($\rho $) and time (t) for the quantum state characterized by n=1. The plot highlights the dynamic behavior of non-classical states under varying squeezing conditions and temporal evolution.}
\label{fig:figure_2}
\end{figure}

\begin{figure}[h]%
\centering
\includegraphics[width=0.9\textwidth]{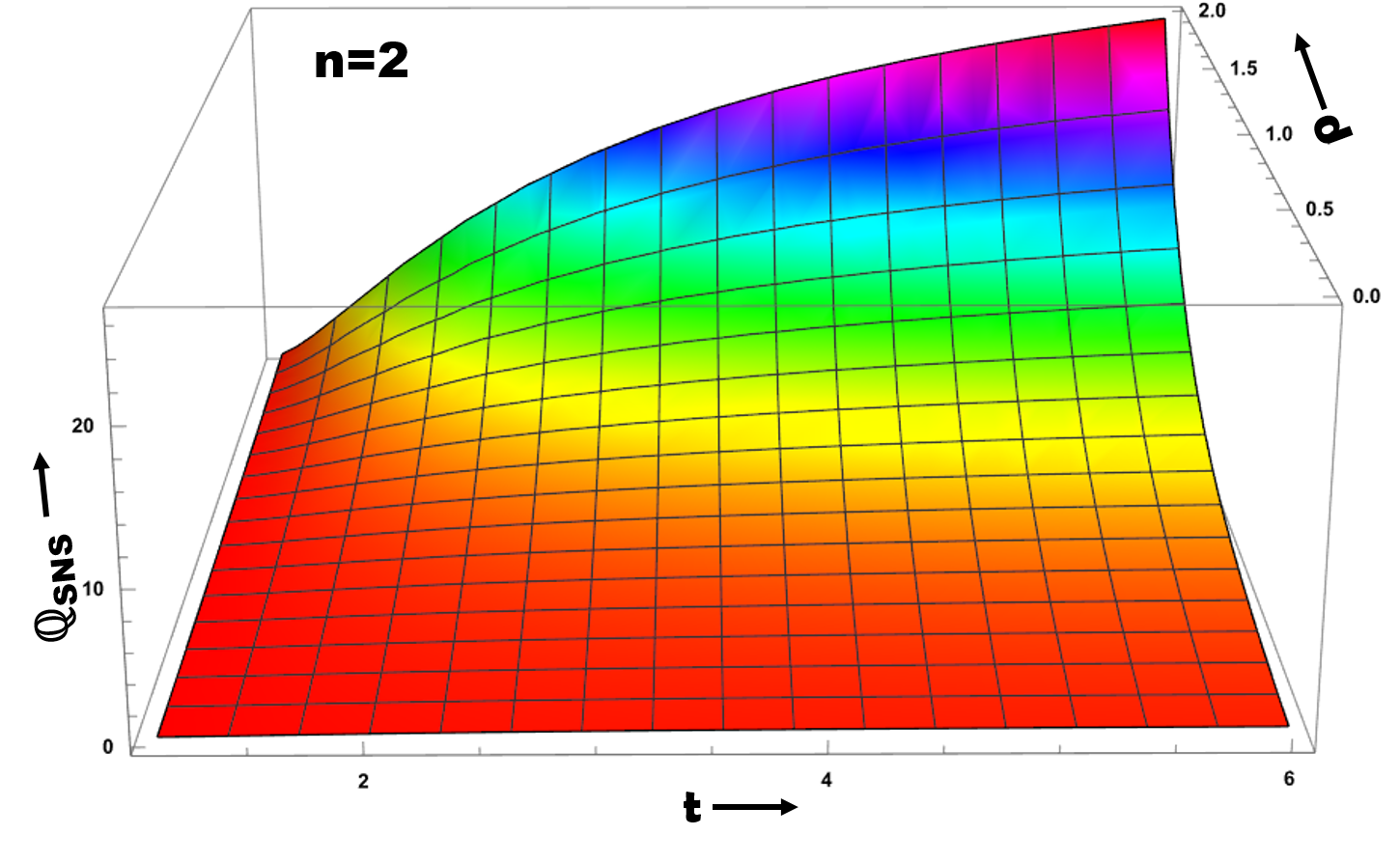}
\caption{3-D representation of the variation in Mandel's $\mathbf{Q_{\text{SNS}}}$ as a function of the squeezing parameter ($\rho $) and time (t) for the quantum state characterized by n=2. The plot highlights the dynamic behavior of non-classical states under varying squeezing conditions and temporal evolution.}
\label{fig:figure_3}
\end{figure}

\begin{figure}[h]%
\centering
\includegraphics[width=0.9\textwidth]{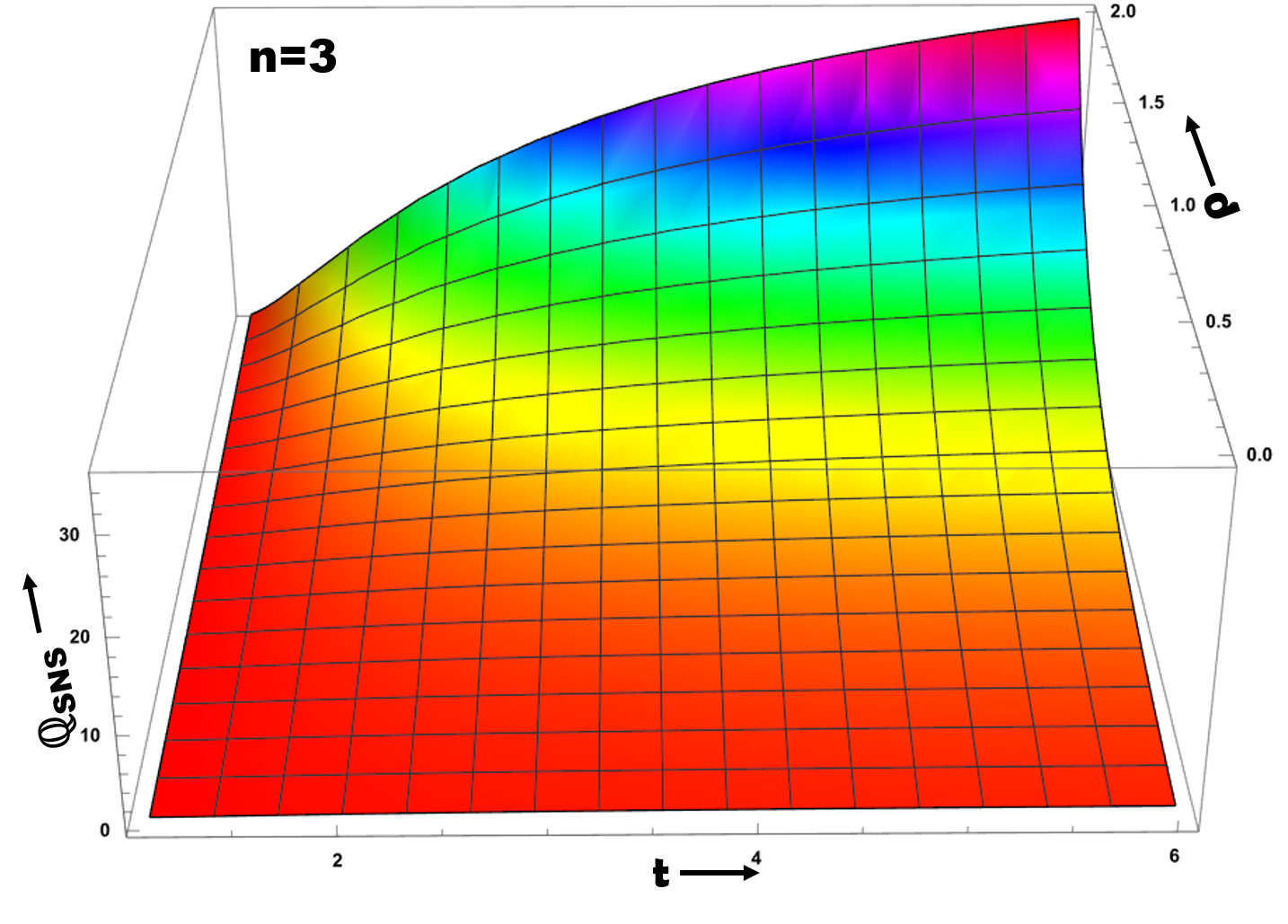}
\caption{3-D representation of the variation in Mandel's $\mathbf{Q_{\text{SNS}}}$ as a function of the squeezing parameter ($\rho $) and time (t) for the quantum state characterized by n=3. The plot highlights the dynamic behavior of non-classical states under varying squeezing conditions and temporal evolution.}
\label{fig:figure_4}
\end{figure}

\begin{figure}[h]%
\centering
\includegraphics[width=0.9\textwidth]{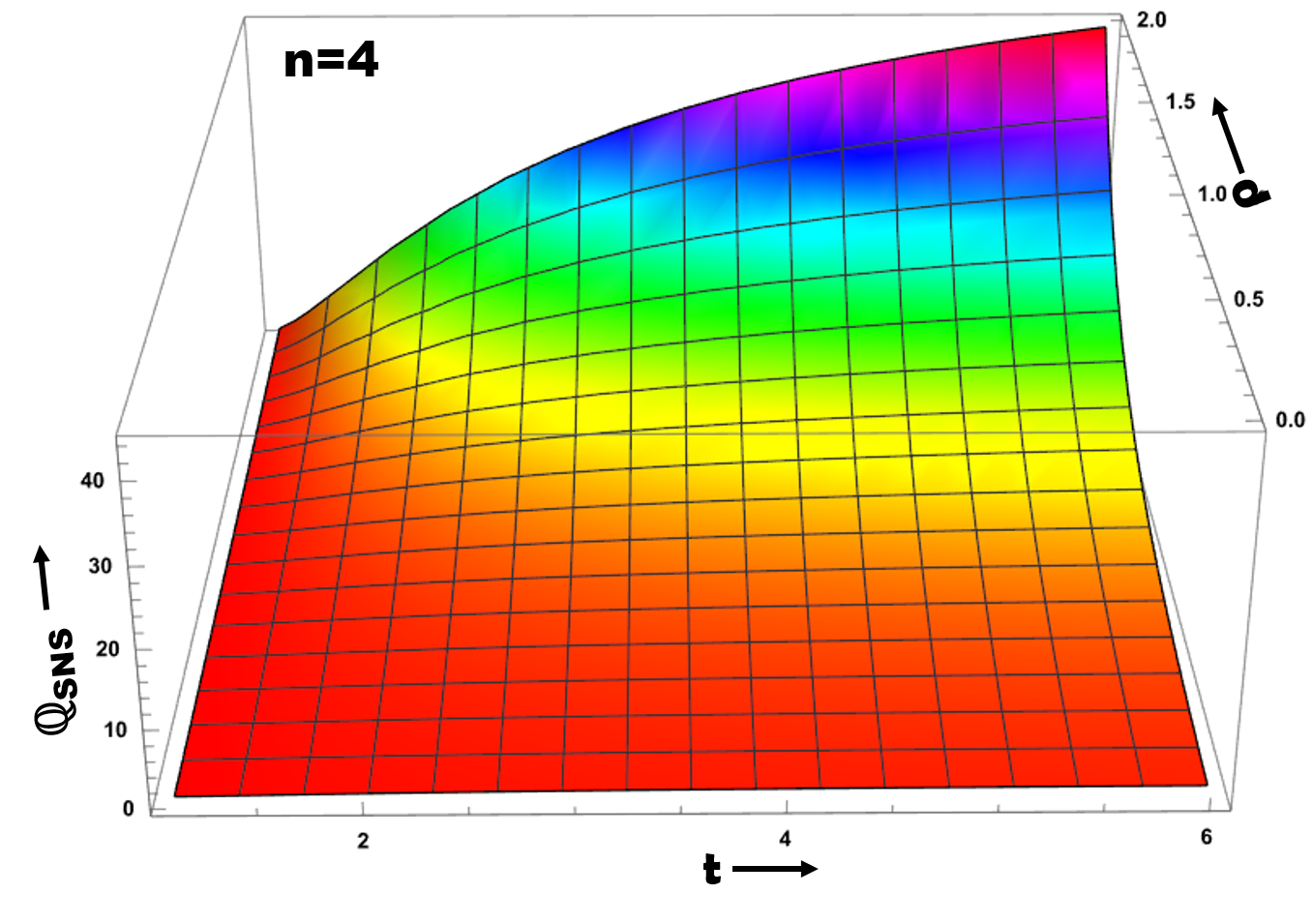}
\caption{3-D representation of the variation in Mandel's $\mathbf{Q_{\text{SNS}}}$ as a function of the squeezing parameter ($\rho $) and time (t) for the quantum state characterized by n=4. The plot highlights the dynamic behavior of non-classical states under varying squeezing conditions and temporal evolution.}
\label{fig:figure_5}
\end{figure}

\begin{figure}[h]%
\centering
\includegraphics[width=0.9\textwidth]{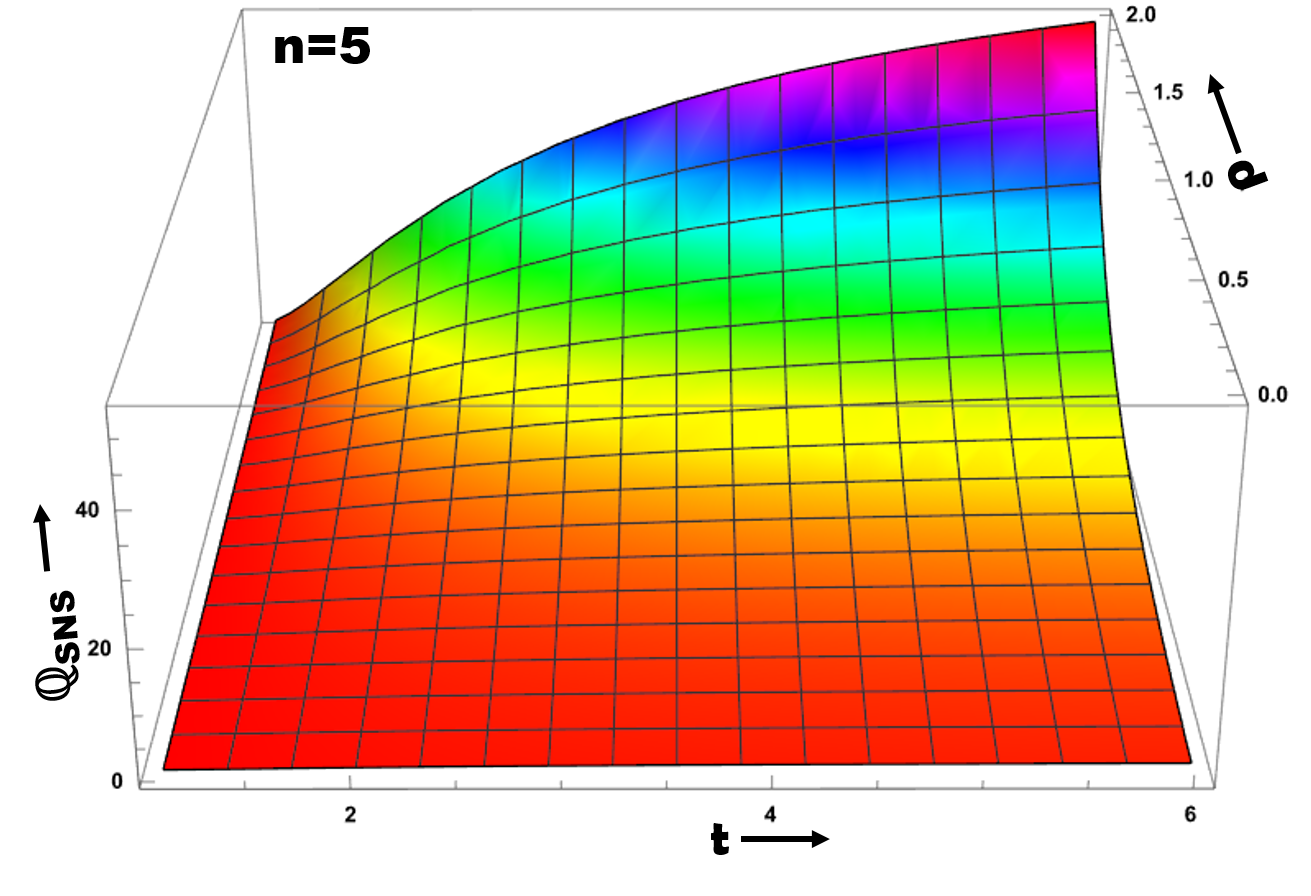}
\caption{3-D representation of the variation in Mandel's $\mathbf{Q_{\text{SNS}}}$ as a function of the squeezing parameter ($\rho $) and time (t) for the quantum state characterized by n=5. The plot highlights the dynamic behavior of non-classical states under varying squeezing conditions and temporal evolution.}
\label{fig:figure_6}
\end{figure}

\begin{figure}[h]%
\centering
\includegraphics[width=0.9\textwidth]{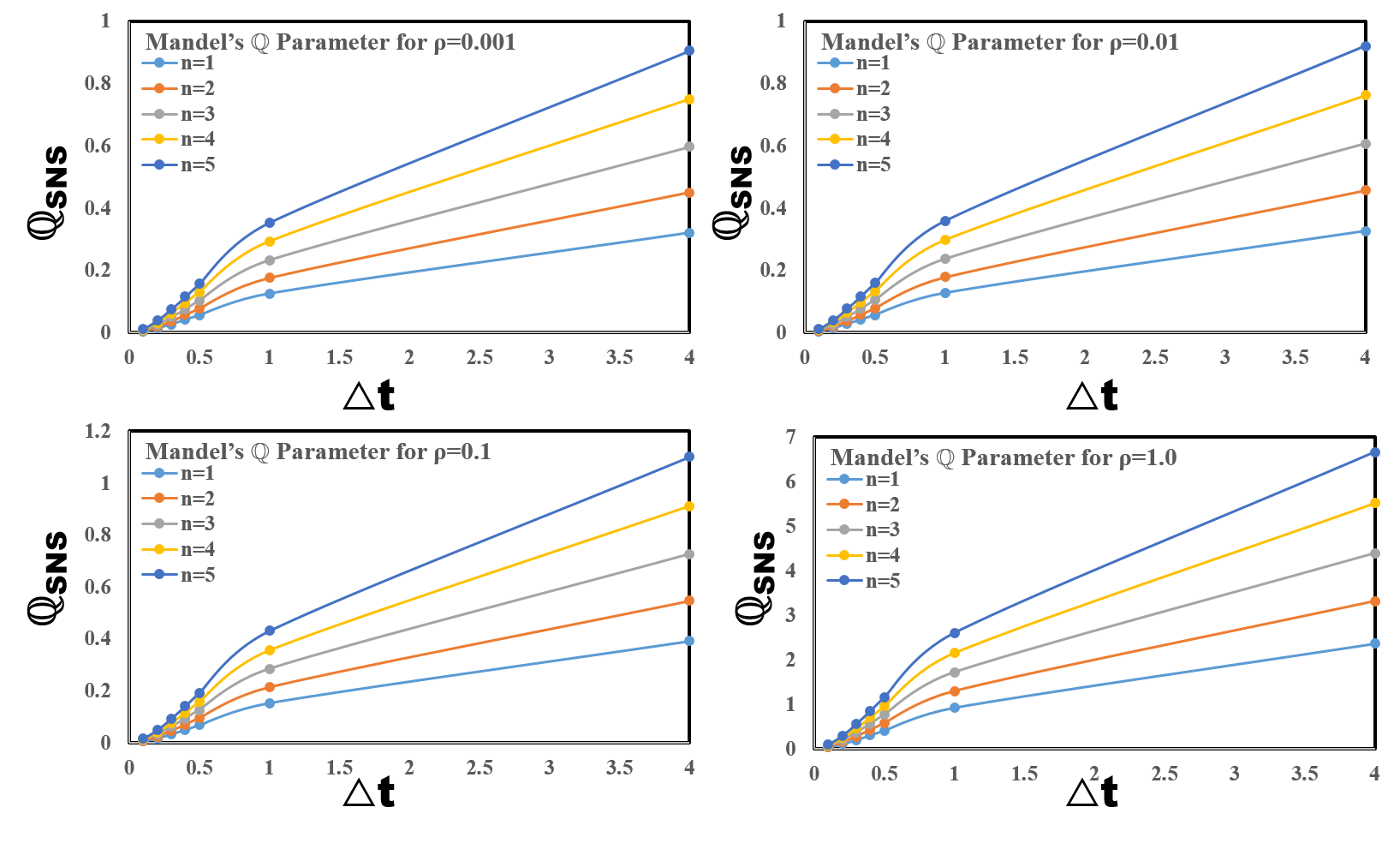}
\caption{Dependence of Mandel's $\mathbf{Q_{\text{SNS}}}$ Parameter on Time ($t$) and Quantum Number ($n$): The plot highlights the non-classical characteristics as influenced by temporal evolution and quantum states.}
\label{fig:figure_7}
\end{figure}

Hence, Cosmological Mandel's $\mathbb{Q}_{\text{SNS}}$ Parameter for using quantum mechanical number state for Squeezed Number State (SNS) is shown by (\ref {5.1.11}). The $\mathbb{Q}_{\text{SNS}}$ is the proportional function of squeezing parameter ($\rho $), number state (n), and inversely proportional to various powers of $\mathbf{t} and \mathbf{t}_0$. The dependency of $\mathbb{Q}_{\text{SNS}}$ is more prominent on number state (n) in comparison to ($\rho $). Using Eqs. (\ref {5.1.11}), calculated values of $\mathbb{Q}_{\text{SNS}}$ for various squeezed number states for n=1, 2, 3, 4, 5 is tabulated in tables \ref{tab:table_1}-\ref{tab:table_5}, for simplicity of evaluation here we are assuming m=$\mathbf{t}_0$=1. Numerical values are calculated for the squeezing parameter ($\rho $) ranging between 0.002 to 2.000, At the same time, time period $\mathbf{t}$ varies from 1.1 to 6 w.r.t to initial time $\mathbf{t}_0$=1, 

 i.e. the $\triangle\mathbf{t}$ is ranging between 0.1 to 5. According to calculations, for $\rho$ ranging between 0.002 to 1.000 variation in Cosmological Mandel's $\mathbb{Q}_{\text{SNS}}$ Parameteris significantly less than that for $\rho$ ranging between 1.000 to 2.000 for all number state and $\triangle\mathbf{t}$ taken into consideration.  For all of these states, the super-Poissonian non-classical nature of the inflaton is demonstrated by the positive values of Cosmological Mandel's $\mathbb{Q}_{\text{SNS}}$. In Eqs. (\ref {5.1.11}) while considering n=0, it  transforms into eq. For Cosmological Mandel's $\mathbb{Q}$ parameter for squeezed vacuum state \cite{ venkataratnam_density_2008, venkataratnam_oscillatory_2010, venkataratnam_behavior_2013} with the nature of evaluation changes from super-Poissonian non-classical nature to sub-Poissonian non-classical nature. we have also depicted variations in Cosmological Mandel's $\mathbb{Q}$ parameter for various squeezed number states with squeezing parameter $\rho $ in Fig. \ref{fig:figure_1}, which exhibits  increasing nature with increasing $\rho $ as well as n.  A 3-D plot between Cosmological Mandel's $\mathbb{Q}$ parameter, $\rho $, and $\mathbf{t}$ for different number state parameters is presented in Fig. (\ref{fig:figure_2}-\ref{fig:figure_6}). While \ref{fig:figure_7} highlights the nonclassical behavior of the inflaton field by showing how the $\mathbb{Q}_{\text{SNS}}$ parameter depends on time and quantum state number for a flat FRW universe. The parameter's sensitivity to quantum states and temporal evolution reflects its utility in characterizing statistical properties of the inflaton within semiclassical gravity frameworks.

\subsection{Cosmological Mandel's $\mathbb{Q}$ Parameter for Coherent Squeezed Number State (CSNS)}

Now to identify the classical or non-classical nature of Coherent Squeezed Number State (CSNS) in FRW Universe, in this section, we will compute Mandel’s $\mathbb{Q}$ Parameter for CSNS using Eq. (\ref{5.6}). Using Eqs. (\ref{3.11}-\ref{3.17}, \ref{4.7}-\ref{4.10}) the values of $\left\langle :\overset{\wedge}{\mathit{e}}^{\dagger2}{\overset{\wedge}{\mathit{e}}}^2:\right\rangle _{\text{CSNS}}$

\begin{align} \label{5.2.1}
\left\langle :\overset{\wedge}{\mathit{e}}^{\dagger2}{\overset{\wedge}{\mathit{e}}}^2:\right\rangle _{\text{CSNS}}=&\biggr[\mathcal{G}^{12}(\mathbf{t})\Phi^2\left(\mathbf{t}\right)\Phi^{*2}\left(\mathbf{t}\right)\dot{\Phi }^2\left(\mathbf{t}_0\right)\dot{\Phi }^{*2}\left(\mathbf{t}_0\right)\nonumber\\
&+\mathcal{G}^{12}(\mathbf{t})\dot{\Phi}^2\left(\mathbf{t}\right)\dot{\Phi}^{*2}\left(\mathbf{t}\right){\Phi }^2\left(\mathbf{t}_0\right){\Phi }^{*2}\left(\mathbf{t}_0\right)\nonumber\\
&-\mathcal{G}^{12}(\mathbf{t})\biggr({\Phi}{^2}\left(\mathbf{t}\right){\Phi}{^*}\left(\mathbf{t}\right)\dot{\Phi}{}\left(\mathbf{t}\right){\Phi}{}\left(\mathbf{t}_0\right)\dot{\Phi}{^*}\left(\mathbf{t}_0\right)\dot{\Phi}{^2}\left(\mathbf{t}_0\right) \nonumber\\
&+ {\Phi}{^2}\left(\mathbf{t}\right){\Phi}{^*}\left(\mathbf{t}\right)\dot{\Phi}{}\left(\mathbf{t}\right){\Phi}{}\left(\mathbf{t}_0\right)\dot{\Phi}^{*2}\left(\mathbf{t}_0\right)\dot{\Phi}{}\left(\mathbf{t}_0\right) \nonumber\\
&+
{\Phi}{}\left(\mathbf{t}\right){\Phi}^{*2}\left(\mathbf{t}\right)\dot{\Phi}{}\left(\mathbf{t}\right){\Phi}{}\left(\mathbf{t}_0\right)\dot{\Phi}{^*}\left(\mathbf{t}_0\right)\dot{\Phi}{^2}\left(\mathbf{t}_0\right) \nonumber\\
&+ {\Phi}{}\left(\mathbf{t}\right){\Phi}^{*2}\left(\mathbf{t}\right)\dot{\Phi}{}\left(\mathbf{t}\right){\Phi}{}\left(\mathbf{t}_0\right)\dot{\Phi}^{*2}\left(\mathbf{t}_0\right)\dot{\Phi}{}\left(\mathbf{t}_0\right)\biggr) \nonumber\\
&-\mathcal{G}^{12}(\mathbf{t})\biggr({\Phi}^{}\left(\mathbf{t}\right)\dot{\Phi}^{*}\left(\mathbf{t}\right)\dot{\Phi}^{2}\left(\mathbf{t}\right){\Phi }^{2}\left(\mathbf{t}_0\right){\Phi }^{*}\left(\mathbf{t}_0\right)\dot{\Phi}^{}\left(\mathbf{t}_0\right) \nonumber\\
& + {\Phi}^{}\left(\mathbf{t}\right)\dot{\Phi}^{*}\left(\mathbf{t}\right)\dot{\Phi}^{2}\left(\mathbf{t}\right){\Phi }^{}\left(\mathbf{t}_0\right){\Phi }^{*2}\left(\mathbf{t}_0\right)\dot{\Phi}^{}\left(\mathbf{t}_0\right) \nonumber\\
& +{\Phi}^{}\left(\mathbf{t}\right)\dot{\Phi}^{*2}\left(\mathbf{t}\right)\dot{\Phi}^{}\left(\mathbf{t}\right){\Phi }^{2}\left(\mathbf{t}_0\right){\Phi }^{*}\left(\mathbf{t}_0\right)\dot{\Phi}^{}\left(\mathbf{t}_0\right) \nonumber\\
& + {\Phi}^{}\left(\mathbf{t}\right)\dot{\Phi}^{*2}\left(\mathbf{t}\right)\dot{\Phi}^{}\left(\mathbf{t}\right){\Phi }^{}\left(\mathbf{t}_0\right){\Phi }^{*2}\left(\mathbf{t}_0\right)\dot{\Phi}^{}\left(\mathbf{t}_0\right)\biggr)\nonumber\\
&+\mathcal{G}^{12}(\mathbf{t})\biggr({\Phi}^{2}\left(\mathbf{t}\right)\dot{\Phi}^{*2}\left(\mathbf{t}\right)\dot{\Phi}^{2}\left(\mathbf{t}_0\right){\Phi }^{*2}\left(\mathbf{t}_0\right) \nonumber\\
& + \dot{\Phi}^{2}\left(\mathbf{t}\right){\Phi }^{*2}\left(\mathbf{t}\right){\Phi}^{2}\left(\mathbf{t}_0\right)\dot{\Phi}^{*2}\left(\mathbf{t}_0\right) \nonumber\\
& +{\Phi}^{}\left(\mathbf{t}\right)\dot{\Phi}^{}\left(\mathbf{t}\right){\Phi}^{*}\left(\mathbf{t}\right)\dot{\Phi}^{*}\left(\mathbf{t}\right){\Phi}^{}\left(\mathbf{t}_0\right)\dot{\Phi}^{}\left(\mathbf{t}_0\right){\Phi}^{*}\left(\mathbf{t}_0\right)\dot{\Phi}^{*}\left(\mathbf{t}_0\right)\nonumber\\
& +{\Phi}^{}\left(\mathbf{t}\right)\dot{\Phi}^{}\left(\mathbf{t}\right){\Phi}^{*}\left(\mathbf{t}\right)\dot{\Phi}^{*}\left(\mathbf{t}\right)\dot{\Phi}^{}\left(\mathbf{t}_0\right){\Phi}^{}\left(\mathbf{t}_0\right)\dot{\Phi}^{*}\left(\mathbf{t}_0\right){\Phi}^{*}\left(\mathbf{t}_0\right)\nonumber\\
& +{\Phi}^{}\left(\mathbf{t}\right)\dot{\Phi}^{}\left(\mathbf{t}\right){\Phi}^{*}\left(\mathbf{t}\right)\dot{\Phi}^{*}\left(\mathbf{t}\right){\Phi}^{2}\left(\mathbf{t}_0\right){\Phi}^{*2}\left(\mathbf{t}_0\right)\nonumber\\
&+{\Phi}^{}\left(\mathbf{t}\right)\dot{\Phi}^{}\left(\mathbf{t}\right){\Phi}^{*}\left(\mathbf{t}\right)\dot{\Phi}^{*}\left(\mathbf{t}\right)\dot{\Phi}^{2}\left(\mathbf{t}_0\right)\dot{\Phi}^{*2}\biggr)\biggr]\biggr\langle n\biggr|\overset{\wedge}W^{\dagger}\mathfrak{D}^{\dagger
}\biggr[6\overset{\wedge
}{\mathit{e}} ^{\dagger2 }\overset{\wedge}{\mathit{e}}^2\nonumber\\
&+12\overset{\wedge }{\mathit{e}} ^{\dagger}\overset{\wedge }{\mathit{e}}+3-6\overset{\wedge }{\mathit{e}} ^{\dagger2 }+4\overset{\wedge }{\mathit{e}} ^{\dagger3}\overset{\wedge }{\mathit{e}}-6\overset{\wedge }{\mathit{e}} ^2+4\overset{\wedge }{\mathit{e}} ^{\dagger }\overset{\wedge }{\mathit{e}}^3+\overset{\wedge }{\mathit{e}} ^{\dagger 4}+\overset{\wedge }{\mathit{e}}^4\biggr]\mathfrak{D}\overset{\wedge}W\biggr|n\biggr\rangle,
\end{align}

using Eqs. (\ref{5.1.2}-\ref{5.1.4}) in Eqs. (\ref{5.2.1}) as

\begin{align} \label{5.2.2}
\left\langle :\overset{\wedge}{\mathit{e}}^{\dagger2}{\overset{\wedge}{\mathit{e}}}^2:\right\rangle _{\text{CSNS}}=&\biggr[\mathcal{G}^{12}(\mathbf{t})\Phi^2\left(\mathbf{t}\right)\Phi^{*2}\left(\mathbf{t}\right)\dot{\Phi }^2\left(\mathbf{t}_0\right)\dot{\Phi }^{*2}\left(\mathbf{t}_0\right)\nonumber\\
&+\mathcal{G}^{12}(\mathbf{t})\dot{\Phi}^2\left(\mathbf{t}\right)\dot{\Phi}^{*2}\left(\mathbf{t}\right){\Phi }^2\left(\mathbf{t}_0\right){\Phi }^{*2}\left(\mathbf{t}_0\right)\nonumber\\
&-\mathbb{X}\mathcal{G}^{12}(\mathbf{t})-\mathbb{Y}\mathcal{G}^{12}(\mathbf{t})+\mathbb{Z}\mathcal{G}^{12}(\mathbf{t})\biggr]\biggr\langle n\biggr|\overset{\wedge}W^{\dagger}\mathfrak{D}^{\dagger
}\biggr[6\overset{\wedge
}{\mathit{e}} ^{\dagger2 }\overset{\wedge}{\mathit{e}}^2+12\overset{\wedge }{\mathit{e}} ^{\dagger}\overset{\wedge }{\mathit{e}}\nonumber\\
&+3-6\overset{\wedge }{\mathit{e}} ^{\dagger2 }+4\overset{\wedge }{\mathit{e}} ^{\dagger3}\overset{\wedge }{\mathit{e}}-6\overset{\wedge }{\mathit{e}} ^2+4\overset{\wedge }{\mathit{e}} ^{\dagger }\overset{\wedge }{\mathit{e}}^3+\overset{\wedge }{\mathit{e}} ^{\dagger 4}+\overset{\wedge }{\mathit{e}}^4\biggr]\mathfrak{D}\overset{\wedge}W\biggr|n\biggr\rangle,
\end{align}

the value of $\left\langle :\overset{\wedge}{\mathit{e}}^{\dagger2}{\overset{\wedge}{\mathit{e}}}^2:\right\rangle _{\text{CSNS}}$ can be further computed using Using Eqs. (\ref{3.11}-\ref{3.17}, \ref{4.7}-\ref{4.10}) in Eq. (\ref{5.2.2})

\begin{align} \label{5.2.3}
\left\langle :\overset{\wedge}{\mathit{e}}^{\dagger2}{\overset{\wedge}{\mathit{e}}}^2:\right\rangle _{\text{CSNS}}=&\left(\frac{1}{16m^{4}\mathbf{t}^4\mathbf{t}_0^4}+\frac{1}{16m^{4}}+\frac{3}{8m^{4}\mathbf{t}^2\mathbf{t}_0^2}-\frac{1}{4m^{4}\mathbf{t}^3\mathbf{t}_0^3}-\frac{1}{4m^{4}\mathbf{t}\mathbf{t}_0}\right)\biggr[3+\Upsilon
^{*4}\nonumber\\
&+\Upsilon ^4+6\Upsilon ^{*2}\Upsilon ^2+12\Upsilon ^*\Upsilon -6\Upsilon ^{*2}-6\Upsilon ^2-4\Upsilon ^{*3}\Upsilon -4\Upsilon ^*\Upsilon ^3\nonumber\\
&+\text{Cosh}^4\rho
(6n^2-6n)+\text{Sinh}^4\rho \left(6n^2+18n+12\right)\nonumber\\
&+\text{Cosh}^2\rho \text{Sinh}^2\rho \left(36n^2+36n+12\right)\nonumber\\
&+\text{Cosh}^3\rho \text{Sinh$\rho
$}\left(24n^2\right)+\text{Cosh$\rho $} \text{Sinh}^3\rho \left(24n^2+48n+24\right)\nonumber\\
&+\text{Cosh$\rho $} \text{Sinh$\rho $}(24n+12)\left\{1+2\Upsilon
^*\Upsilon -\Upsilon ^{*2}-\Upsilon ^2\right\}\nonumber\\
&+\text{Sinh}^2\rho (12n+12)\left\{1+2\Upsilon ^*\Upsilon -\Upsilon ^{*2}-\Upsilon ^2\right\}\nonumber\\
&+\text{Cosh}^2\rho(12n)\left\{1+2\Upsilon ^*\Upsilon
-\Upsilon ^{*2}-\Upsilon ^2\right\}\biggr].
\end{align}

using Eqs. (\ref{3.11}-\ref{3.17}, \ref{4.7}-\ref{4.10}) the values of $\left\langle :\overset{\wedge}{\mathit{e}}^{\dagger}{\overset{\wedge}{\mathit{e}}}:\right\rangle _{\text{CSNS}}$ 

\begin{align} \label{5.2.4}
\left\langle :\overset{\wedge}{\mathit{e}}^{\dagger}{\overset{\wedge}{\mathit{e}}}:\right\rangle _{\text{CSNS}}=&\biggr[\mathcal{G}^{6}(\mathbf{t}){\Phi}\left(\mathbf{t}\right){\Phi}^*\left(\mathbf{t}\right)\dot{\Phi }\left(\mathbf{t}_0\right)\dot{\Phi}\left(\mathbf{t}_0\right)\nonumber\\
&+\mathcal{G}^{6}(\mathbf{t})\dot{\Phi}\left(\mathbf{t}\right)\dot{\Phi}^*\left(\mathbf{t}\right){\Phi }\left(\mathbf{t}_0\right){\Phi}\left(\mathbf{t}_0\right)\nonumber\\
&-\mathcal{G}^{6}(\mathbf{t})\dot{\Phi}\left(\mathbf{t}\right){\Phi}^*\left(\mathbf{t}\right){\Phi }\left(\mathbf{t}_0\right)\dot{\Phi}^*\left(\mathbf{t}_0\right)\nonumber\\
&-\mathcal{G}^{6}(\mathbf{t}){\Phi}\left(\mathbf{t}\right)\dot{\Phi}^*\left(\mathbf{t}\right){\Phi }^*\left(\mathbf{t}_0\right)\dot{\Phi}\left(\mathbf{t}_0\right)\biggr]\biggr\langle n\biggr|\overset{\wedge}W^{\dagger}\mathfrak{D}^{\dagger
}\biggr[2\overset{\wedge }{\mathit{e}} ^{\dagger }\overset{\wedge }{\mathit{e}}\nonumber\\
&+1-\overset{\wedge }{\mathit{e}} ^{\dagger2 }-\overset{\wedge }{\mathit{e}} ^2\biggr]\mathfrak{D}\overset{\wedge}W\biggr|n\biggr\rangle,
\end{align}

the value of $\left\langle :\overset{\wedge}{\mathit{e}}^{\dagger}{\overset{\wedge}{\mathit{e}}}:\right\rangle _{\text{CSNS}}$ can be further computed using Using Eqs. (\ref{3.11}-\ref{3.17}, \ref{4.7}-\ref{4.10}) in Eq. (\ref{5.2.4})

\begin{align} \label{5.2.5}
\left\langle :\overset{\wedge}{\mathit{e}}^{\dagger}{\overset{\wedge}{\mathit{e}}}:\right\rangle _{\text{CSNS}}=&\left(\frac{1}{4m^{2}\mathbf{t}^2\mathbf{t}_0^2}+\frac{1}{4m^{2}}-\frac{1}{2m^{2}\mathbf{t}\mathbf{t}_0}\right)[\text{Cosh}^2\rho (2n)+\text{Sinh}^2\rho
(2n+2)\nonumber\\
&+\text{Cosh$\rho $} \text{Sinh$\rho $}(4n+2)+1+\Upsilon ^*\Upsilon -\Upsilon ^{*2}-\Upsilon ^2].
\end{align}

Using Eqs. (\ref{3.11}-\ref{3.17}, \ref{4.7}-\ref{4.10}) the values of $\left\langle :\overset{\wedge}{\mathit{e}}^{\dagger}{\overset{\wedge}{\mathit{e}}}:\right\rangle^2_{\text{CSNS}}$ 

\begin{align} \label{5.2.6}
\left\langle :\overset{\wedge}{\mathit{e}}^{\dagger}{\overset{\wedge}{\mathit{e}}}:\right\rangle^2_{\text{SNS}}=&\biggr[\mathcal{G}^{6}(\mathbf{t}){\Phi}\left(\mathbf{t}\right){\Phi}^*\left(\mathbf{t}\right)\dot{\Phi }\left(\mathbf{t}_0\right)\dot{\Phi}\left(\mathbf{t}_0\right)\nonumber\\
&+\mathcal{G}^{6}(\mathbf{t})\dot{\Phi}\left(\mathbf{t}\right)\dot{\Phi}^*\left(\mathbf{t}\right){\Phi }\left(\mathbf{t}_0\right){\Phi}\left(\mathbf{t}_0\right)\nonumber\\
&-\mathcal{G}^{6}(\mathbf{t})\dot{\Phi}\left(\mathbf{t}\right){\Phi}^*\left(\mathbf{t}\right){\Phi }\left(\mathbf{t}_0\right)\dot{\Phi}^*\left(\mathbf{t}_0\right)\nonumber\\
&-\mathcal{G}^{6}(\mathbf{t}){\Phi}\left(\mathbf{t}\right)\dot{\Phi}^*\left(\mathbf{t}\right){\Phi }^*\left(\mathbf{t}_0\right)\dot{\Phi}\left(\mathbf{t}_0\right)\biggr]^2\biggr\langle n\biggr|\overset{\wedge}W^{\dagger}\mathfrak{D}^{\dagger
}\biggr[2\overset{\wedge }{\mathit{e}} ^{\dagger }\overset{\wedge }{\mathit{e}}\nonumber\\
&+1-\overset{\wedge }{\mathit{e}} ^{\dagger2 }-\overset{\wedge }{\mathit{e}} ^2\biggr]^2\mathfrak{D}\overset{\wedge}W\biggr|n\biggr\rangle,
\end{align}

the value of $\left\langle :\overset{\wedge}{\mathit{e}}^{\dagger}{\overset{\wedge}{\mathit{e}}}:\right\rangle^2_{\text{CSNS}}$ can be further computed using Using Eqs. (\ref{3.11}-\ref{3.17}, \ref{4.7}-\ref{4.10}) in Eq. (\ref{5.2.6})

\begin{align} \label{5.2.7}
\left\langle :\overset{\wedge}{\mathit{e}}^{\dagger}{\overset{\wedge}{\mathit{e}}}:\right\rangle^2_{\text{CSNS}}=&\left(\frac{1}{16m^{4}\mathbf{t}^4\mathbf{t}_0^4}+\frac{1}{16m^{4}}+\frac{3}{8m^{4}\mathbf{t}^2\mathbf{t}_0^2}-\frac{1}{4m^{4}\mathbf{t}^3\mathbf{t}_0^3}-\frac{1}{4m^{4}\mathbf{t}\mathbf{t}_0}\right)\biggr[1\nonumber\\
&+\Upsilon
^{*4}+\Upsilon ^4+3\Upsilon ^{*2}\Upsilon ^2\nonumber\\
&+2\Upsilon ^*\Upsilon -2\Upsilon ^{*2}-2\Upsilon ^2-2\Upsilon ^{*3}\Upsilon -2\Upsilon ^*\Upsilon ^3\nonumber\\
&+\text{Cosh}^4\rho
\left(4n^2\right)+\text{Sinh}^4\rho \left(4n^2+8n+4\right)\nonumber\\
&+\text{Cosh}^2\rho\text{Sinh}^2\rho \left(24n^2+24n+4\right)+\text{Cosh}^3\rho \text{Sinh$\rho
$}(16n^2\nonumber\\
&+8n)+\text{Cosh$\rho $} \text{Sinh}^3\rho \left(16n^2+24n+8\right)\nonumber\\
&+\text{Cosh$\rho $} \text{Sinh$\rho $}(8n+4)\left\{1+\Upsilon
^*\Upsilon -\Upsilon ^{*2}-\Upsilon ^2\right\}\nonumber\\
&+\text{Sinh}^2\rho (4n+4)\left\{1+\Upsilon ^*\Upsilon -\Upsilon ^{*2}-\Upsilon ^2\right\}\nonumber\\
&+\text{Cosh}^2\rho
(4n)\left\{1+\Upsilon ^*\Upsilon -\Upsilon ^{*2}-\Upsilon ^2\right\}\biggr].
\end{align}

Substituting the values of $\left\langle :\overset{\wedge}{\mathit{e}}^{\dagger2}{\overset{\wedge}{\mathit{e}}}^2:\right\rangle _{\text{CSNS}}$, $\left\langle :\overset{\wedge}{\mathit{e}}^{\dagger}{\overset{\wedge}{\mathit{e}}}:\right\rangle _{\text{CSNS}}$ and $\left\langle :\overset{\wedge}{\mathit{e}}^{\dagger}{\overset{\wedge}{\mathit{e}}}:\right\rangle^2_{\text{CSNS}}$ in Eq. (\ref{5.6}), Cosmological Mandel's $\mathbb{Q}$ Parameter for Coherent squeezed number state is

\begin{align} \label{5.2.8}
\mathbb{Q}_{\text{CSNS}}=&\left(\frac{1}{4m^{2}\mathbf{t}^2\mathbf{t}_0^2}+\frac{1}{4m^{2}}-\frac{1}{2m^{2}\mathbf{t}\mathbf{t}_0}\right)\biggr[2+3\Upsilon ^{*2}\Upsilon ^2+10\Upsilon
^*\Upsilon -4\Upsilon ^{*2}\nonumber\\
&-4\Upsilon ^2-2\Upsilon ^{*3}\Upsilon -2\Upsilon ^*\Upsilon ^3+\text{Cosh}^4\rho (2n^2-6n)\nonumber\\
&+\text{Sinh}^4\rho \left(2n^2+10n+8\right)+\text{Cosh}^2\rho
 \text{Sinh}^2\rho \left(12n^2+12n+8\right)\nonumber\\
&+\text{Cosh}^3\rho\text{Sinh$\rho $}\left(8n^2-8n\right)+\text{Cosh$\rho $}\text{Sinh}^3\rho \left(8n^2+24n+16\right)\nonumber\\
&+\text{Cosh$\rho
$}\text{Sinh$\rho $}(16n+8)\left\{1-\varUpsilon ^{*2}-\varUpsilon ^2\right\}+\text{Cosh$\rho $} \text{Sinh$\rho $}(40n\nonumber\\
&+20)\left\{\varUpsilon ^*\varUpsilon
\right\}+\text{Sinh}^2\rho (8n+8)\left\{1-\varUpsilon^{*2}-\varUpsilon ^2\right\}\nonumber\\
&+\text{Sinh}^2\rho (20n+20)\left\{\varUpsilon ^*\varUpsilon \right\}+\text{Cosh}^2\rho
(8n)\left\{1-\varUpsilon ^{*2}-\varUpsilon ^2\right\}\nonumber\\
&+\text{Cosh}^2\rho (20n)\left\{\varUpsilon ^*\varUpsilon \right\}\biggr]{/}\biggr[\text{Cosh}^2\rho (2n)+\text{Sinh}^2\rho
(2n+2)\nonumber\\
&+\text{Cosh$\rho $} \text{Sinh$\rho $}(4n+2)+1+\Upsilon ^*\Upsilon -\Upsilon ^{*2}-\Upsilon ^2\biggr]
\end{align}

\begin{table}[h]\label{table 6}
\caption{Numerical values of Mandel's $\mathbf{Q}$ Parameter while n=1, $\mathbf{t}_0=1$ for various values of $\rho$ and $\triangle\mathbf{t}$}
\label{tab:table_6}
\begin{tabular}{@{}llllllll@{}}
\toprule
$\rho$ & $\triangle\mathbf{t}=0.1 $  & $\triangle\mathbf{t}=0.2 $ & $\triangle\mathbf{t}=0.3 $ & $ \triangle\mathbf{t}=0.4 $ & $ \triangle\mathbf{t}=0.5$ & $\triangle\mathbf{t}=1 $ & $\triangle\mathbf{t}=5 $\\
\midrule
0.002 & 0.0114 & 0.0382 & 0.0733 & 0.1123 & 0.1529 & 0.3439 & 0.8805 \\ 
0.004 & 0.0114 & 0.0382 & 0.0733 & 0.1124 & 0.1529 & 0.3441 & 0.8810 \\ 
0.006 & 0.0114 & 0.0383 & 0.0733 & 0.1124 & 0.1530 & 0.3443 & 0.8815 \\ 
0.008 & 0.0114 & 0.0383 & 0.0734 & 0.1125 & 0.1531 & 0.3445 & 0.8820 \\ 
0.010 & 0.0114 & 0.0383 & 0.0734 & 0.1126 & 0.1532 & 0.3447 & 0.8825 \\ 
0.020 & 0.0114 & 0.0384 & 0.0737 & 0.1129 & 0.1537 & 0.3458 & 0.8854 \\ 
0.040 & 0.0115 & 0.0387 & 0.0742 & 0.1138 & 0.1548 & 0.3484 & 0.8918 \\ 
0.060 & 0.0116 & 0.0390 & 0.0748 & 0.1147 & 0.1561 & 0.3513 & 0.8994 \\ 
0.080 & 0.0117 & 0.0394 & 0.0756 & 0.1158 & 0.1576 & 0.3547 & 0.9080 \\ 
0.100 & 0.0118 & 0.0398 & 0.0764 & 0.1170 & 0.1593 & 0.3584 & 0.9176 \\ 
0.200 & 0.0127 & 0.0426 & 0.0816 & 0.1251 & 0.1703 & 0.3831 & 0.9808 \\ 
0.400 & 0.0153 & 0.0515 & 0.0987 & 0.1513 & 0.2059 & 0.4632 & 1.1858 \\ 
0.600 & 0.0196 & 0.0659 & 0.1264 & 0.1937 & 0.2637 & 0.5933 & 1.5189 \\ 
0.800 & 0.0262 & 0.0881 & 0.1690 & 0.2590 & 0.3526 & 0.7933 & 2.0309 \\ 
1.000 & 0.0362 & 0.1217 & 0.2333 & 0.3576 & 0.4868 & 1.0952 & 2.8038 \\ 
1.200 & 0.0512 & 0.1720 & 0.3297 & 0.5054 & 0.6879 & 1.5478 & 3.9624 \\ 
1.400 & 0.0735 & 0.2472 & 0.4738 & 0.7263 & 0.9886 & 2.2244 & 5.6945 \\ 
1.600 & 0.1069 & 0.3594 & 0.6890 & 1.0562 & 1.4376 & 3.2347 & 8.2808 \\ 
1.800 & 0.1568 & 0.5269 & 1.0102 & 1.5485 & 2.1077 & 4.7424 & 12.1406 \\ 
2.000 & 0.2311 & 0.7769 & 1.4894 & 2.2831 & 3.1076 & 6.9921 & 17.8997 \\ 

\botrule
\end{tabular}
\end{table}

\begin{table}[h]\label{table 7}
\caption{Numerical values of Mandel's $\mathbf{Q}$ Parameter while n=2, $\mathbf{t}_0=1$ for various values of $\rho$ and $\triangle\mathbf{t}$}
\label{tab:table_7}
\begin{tabular}{@{}llllllll@{}}
\toprule
$\rho$ & $\triangle\mathbf{t}=0.1 $  & $\triangle\mathbf{t}=0.2 $ & $\triangle\mathbf{t}=0.3 $ & $ \triangle\mathbf{t}=0.4 $ & $ \triangle\mathbf{t}=0.5$ & $\triangle\mathbf{t}=1 $ & $\triangle\mathbf{t}=5 $\\
\midrule
0.002 & 0.0119 & 0.0400 & 0.0767 & 0.1175 & 0.1600 & 0.3600 & 0.9215 \\
0.004 & 0.0119 & 0.0401 & 0.0768 & 0.1177 & 0.1602 & 0.3605 & 0.9230 \\
0.006 & 0.0119 & 0.0401 & 0.0769 & 0.1179 & 0.1605 & 0.3611 & 0.9245 \\
0.008 & 0.0120 & 0.0402 & 0.0771 & 0.1181 & 0.1608 & 0.3617 & 0.9260 \\
0.010 & 0.0120 & 0.0403 & 0.0772 & 0.1183 & 0.1610 & 0.3623 & 0.9275 \\
0.020 & 0.0121 & 0.0406 & 0.0778 & 0.1193 & 0.1624 & 0.3653 & 0.9353 \\
0.040 & 0.0123 & 0.0413 & 0.0792 & 0.1214 & 0.1652 & 0.3717 & 0.9514 \\
0.060 & 0.0125 & 0.0420 & 0.0806 & 0.1235 & 0.1681 & 0.3783 & 0.9685 \\
0.080 & 0.0127 & 0.0428 & 0.0821 & 0.1258 & 0.1713 & 0.3854 & 0.9866 \\
0.100 & 0.0130 & 0.0436 & 0.0837 & 0.1283 & 0.1746 & 0.3928 & 1.0057 \\
0.200 & 0.0144 & 0.0485 & 0.0929 & 0.1424 & 0.1939 & 0.4362 & 1.1166 \\
0.400 & 0.0185 & 0.0621 & 0.1191 & 0.1826 & 0.2485 & 0.5591 & 1.4313 \\
0.600 & 0.0247 & 0.0830 & 0.1592 & 0.2440 & 0.3321 & 0.7473 & 1.9130 \\
0.800 & 0.0341 & 0.1145 & 0.2196 & 0.3366 & 0.4582 & 1.0309 & 2.6391 \\
1.000 & 0.0481 & 0.1618 & 0.3101 & 0.4754 & 0.6470 & 1.4558 & 3.7268 \\
1.200 & 0.0691 & 0.2323 & 0.4454 & 0.6827 & 0.9293 & 2.0909 & 5.3526 \\
1.400 & 0.1005 & 0.3377 & 0.6474 & 0.9923 & 1.3507 & 3.0390 & 7.7799 \\
1.600 & 0.1472 & 0.4949 & 0.9488 & 1.4544 & 1.9796 & 4.4540 & 11.4022 \\
1.800 & 0.2170 & 0.7295 & 1.3985 & 2.1438 & 2.9179 & 6.5653 & 16.8070 \\
2.000 & 0.3212 & 1.0795 & 2.0695 & 3.1723 & 4.3178 & 9.7151 & 24.8706 \\
\botrule
\end{tabular}
\end{table}

\begin{table}[h]\label{table 8}
\caption{Numerical values of Mandel's $\mathbf{Q}$ Parameter while n=3, $\mathbf{t}_0=1$ for various values of $\rho$ and $\triangle\mathbf{t}$}
\label{tab:table_8}
\begin{tabular}{@{}llllllll@{}}
\toprule
$\rho$ & $\triangle\mathbf{t}=0.1 $  & $\triangle\mathbf{t}=0.2 $ & $\triangle\mathbf{t}=0.3 $ & $ \triangle\mathbf{t}=0.4 $ & $ \triangle\mathbf{t}=0.5$ & $\triangle\mathbf{t}=1 $ & $\triangle\mathbf{t}=5 $\\
\midrule
0.002 & 0.0135 & 0.0452 & 0.0867 & 0.1329 & 0.1809 & 0.4071 & 1.0422 \\ 
0.004 & 0.0135 & 0.0453 & 0.0869 & 0.1332 & 0.1813 & 0.4080 & 1.0444 \\ 
0.006 & 0.0135 & 0.0454 & 0.0871 & 0.1335 & 0.1817 & 0.4088 & 1.0466 \\ 
0.008 & 0.0135 & 0.0455 & 0.0873 & 0.1338 & 0.1821 & 0.4097 & 1.0488 \\ 
0.010 & 0.0136 & 0.0456 & 0.0875 & 0.1341 & 0.1825 & 0.4106 & 1.0511 \\ 
0.020 & 0.0137 & 0.0461 & 0.0884 & 0.1355 & 0.1844 & 0.4150 & 1.0624 \\ 
0.040 & 0.0140 & 0.0471 & 0.0904 & 0.1385 & 0.1885 & 0.4242 & 1.0859 \\ 
0.060 & 0.0143 & 0.0482 & 0.0924 & 0.1416 & 0.1928 & 0.4338 & 1.1105 \\ 
0.080 & 0.0147 & 0.0493 & 0.0945 & 0.1449 & 0.1973 & 0.4438 & 1.1362 \\ 
0.100 & 0.0150 & 0.0505 & 0.0968 & 0.1484 & 0.2019 & 0.4544 & 1.1632 \\ 
0.200 & 0.0170 & 0.0572 & 0.1096 & 0.1680 & 0.2287 & 0.5146 & 1.3174 \\ 
0.400 & 0.0225 & 0.0757 & 0.1451 & 0.2225 & 0.3028 & 0.6814 & 1.7443 \\ 
0.600 & 0.0308 & 0.1037 & 0.1988 & 0.3047 & 0.4148 & 0.9332 & 2.3890 \\ 
0.800 & 0.0433 & 0.1456 & 0.2792 & 0.4280 & 0.5826 & 1.3108 & 3.3557 \\ 
1.000 & 0.0620 & 0.2084 & 0.3995 & 0.6124 & 0.8335 & 1.8754 & 4.8009 \\ 
1.200 & 0.0899 & 0.3020 & 0.5791 & 0.8876 & 1.2082 & 2.7184 & 6.9590 \\ 
1.400 & 0.1315 & 0.4418 & 0.8471 & 1.2985 & 1.7673 & 3.9765 & 10.1799 \\ 
1.600 & 0.1935 & 0.6504 & 1.2470 & 1.9114 & 2.6017 & 5.8538 & 14.9857 \\ 
1.800 & 0.2861 & 0.9616 & 1.8436 & 2.8260 & 3.8465 & 8.6546 & 22.1557 \\ 
2.000 & 0.4242 & 1.4259 & 2.7337 & 4.1904 & 5.7036 & 12.833 & 32.8525 \\ 
\botrule
\end{tabular}
\end{table}

\begin{table}[h]\label{table 9}
\caption{Numerical values of Mandel's $\mathbf{Q}$ Parameter while n=4, $\mathbf{t}_0=1$ for various values of $\rho$ and $\triangle\mathbf{t}$}
\label{tab:table_9}
\begin{tabular}{@{}llllllll@{}}
\toprule
$\rho$ & $\triangle\mathbf{t}=0.1 $  & $\triangle\mathbf{t}=0.2 $ & $\triangle\mathbf{t}=0.3 $ & $ \triangle\mathbf{t}=0.4 $ & $ \triangle\mathbf{t}=0.5$ & $\triangle\mathbf{t}=1 $ & $\triangle\mathbf{t}=5 $\\
\midrule
0.002 & 0.0153 & 0.0513 & 0.0984 & 0.1509 & 0.2054 & 0.4621 & 1.1829 \\ 
0.004 & 0.0153 & 0.0515 & 0.0987 & 0.1512 & 0.2059 & 0.4632 & 1.1857 \\ 
0.006 & 0.0153 & 0.0516 & 0.0989 & 0.1516 & 0.2064 & 0.4643 & 1.1886 \\ 
0.008 & 0.0154 & 0.0517 & 0.0991 & 0.1520 & 0.2069 & 0.4654 & 1.1915 \\ 
0.010 & 0.0154 & 0.0518 & 0.0994 & 0.1523 & 0.2074 & 0.4666 & 1.1944 \\ 
0.020 & 0.0156 & 0.0525 & 0.1006 & 0.1542 & 0.2099 & 0.4723 & 1.2091 \\ 
0.040 & 0.0160 & 0.0538 & 0.1031 & 0.1581 & 0.2152 & 0.4842 & 1.2396 \\ 
0.060 & 0.0164 & 0.0552 & 0.1058 & 0.1622 & 0.2207 & 0.4966 & 1.2714 \\ 
0.080 & 0.0168 & 0.0566 & 0.1086 & 0.1664 & 0.2265 & 0.5096 & 1.3046 \\ 
0.100 & 0.0180 & 0.0581 & 0.1114 & 0.1708 & 0.2325 & 0.5231 & 1.3393 \\ 
0.200 & 0.0198 & 0.0667 & 0.1278 & 0.1960 & 0.2667 & 0.6002 & 1.5364 \\ 
0.400 & 0.0268 & 0.0902 & 0.1729 & 0.2650 & 0.3607 & 0.8115 & 2.0775 \\ 
0.600 & 0.0373 & 0.1255 & 0.2405 & 0.3687 & 0.5018 & 1.1291 & 2.8904 \\ 
0.800 & 0.0530 & 0.1782 & 0.3417 & 0.5238 & 0.7130 & 1.6042 & 4.1068 \\ 
1.000 & 0.0765 & 0.2571 & 0.4929 & 0.7556 & 1.0284 & 2.3140 & 5.9238 \\ 
1.200 & 0.1115 & 0.3748 & 0.7186 & 1.1015 & 1.4993 & 3.3735 & 8.6361 \\ 
1.400 & 0.1638 & 0.5505 & 1.0554 & 1.6178 & 2.2020 & 4.9544 & 12.6833 \\ 
1.600 & 0.2418 & 0.8126 & 1.5578 & 2.3880 & 3.2503 & 7.3132 & 18.7218 \\ 
1.800 & 0.3581 & 1.2036 & 2.3074 & 3.5370 & 4.8143 & 10.8322 & 27.7304 \\ 
2.000 & 0.5316 & 1.7869 & 3.4258 & 5.2513 & 7.1476 & 16.0821 & 41.1701 \\ 
\botrule
\end{tabular}
\end{table}

\begin{table}[h]\label{table 10}
\caption{Numerical values of Mandel's $\mathbf{Q}$ Parameter while n=5, $\mathbf{t}_0=1$ for various values of $\rho$ and $\triangle\mathbf{t}$}
\label{tab:table_10}
\begin{tabular}{@{}llllllll@{}}
\toprule
$\rho$ & $\triangle\mathbf{t}=0.1 $  & $\triangle\mathbf{t}=0.2 $ & $\triangle\mathbf{t}=0.3 $ & $ \triangle\mathbf{t}=0.4 $ & $ \triangle\mathbf{t}=0.5$ & $\triangle\mathbf{t}=1 $ & $\triangle\mathbf{t}=5 $\\
\midrule
0.002 & 0.0172 & 0.0578 & 0.1108 & 0.1698 & 0.2312 & 0.5201 & 1.3315 \\ 
0.004 & 0.0172 & 0.0579 & 0.1111 & 0.1703 & 0.2318 & 0.5215 & 1.3350 \\ 
0.006 & 0.0173 & 0.0581 & 0.1114 & 0.1707 & 0.2324 & 0.5229 & 1.3386 \\ 
0.008 & 0.0173 & 0.0583 & 0.1117 & 0.1712 & 0.2330 & 0.5243 & 1.3421 \\ 
0.010 & 0.0174 & 0.0584 & 0.1120 & 0.1716 & 0.2336 & 0.5257 & 1.3457 \\ 
0.020 & 0.0176 & 0.0592 & 0.1135 & 0.1739 & 0.2368 & 0.5327 & 1.3638 \\ 
0.040 & 0.0181 & 0.0608 & 0.1166 & 0.1787 & 0.2432 & 0.5473 & 1.4011 \\ 
0.060 & 0.0186 & 0.0625 & 0.1198 & 0.1837 & 0.2500 & 0.5625 & 1.4400 \\ 
0.080 & 0.0191 & 0.0643 & 0.1232 & 0.1889 & 0.2570 & 0.5784 & 1.4806 \\ 
0.100 & 0.0197 & 0.0661 & 0.1267 & 0.1943 & 0.2644 & 0.5949 & 1.5229 \\ 
0.200 & 0.0228 & 0.0765 & 0.1467 & 0.2249 & 0.3061 & 0.6887 & 1.7630 \\ 
0.400 & 0.0312 & 0.1050 & 0.2013 & 0.3086 & 0.4200 & 0.9451 & 2.4193 \\ 
0.600 & 0.0439 & 0.1477 & 0.2832 & 0.4340 & 0.5908 & 1.3293 & 3.4029 \\ 
0.800 & 0.0629 & 0.2115 & 0.4055 & 0.6216 & 0.8461 & 1.9036 & 4.8732 \\ 
1.000 & 0.0913 & 0.3068 & 0.5882 & 0.9016 & 1.2272 & 2.7612 & 7.0686 \\ 
1.200 & 0.1336 & 0.4490 & 0.8608 & 1.3195 & 1.7960 & 4.0410 & 10.3449 \\ 
1.400 & 0.1967 & 0.6612 & 1.2676 & 1.9430 & 2.6447 & 5.9506 & 15.2334 \\ 
1.600 & 0.2909 & 0.9777 & 1.8744 & 2.8733 & 3.9109 & 8.7995 & 22.5267 \\ 
1.800 & 0.4314 & 1.4500 & 2.7798 & 4.2612 & 5.7999 & 13.0498 & 33.4075 \\ 
2.000 & 0.6410 & 2.1545 & 4.1305 & 6.3316 & 8.6181 & 19.3906 & 49.6400 \\ 
\botrule
\end{tabular}
\end{table}

\begin{figure}[h]%
\centering
\includegraphics[width=0.9\textwidth]{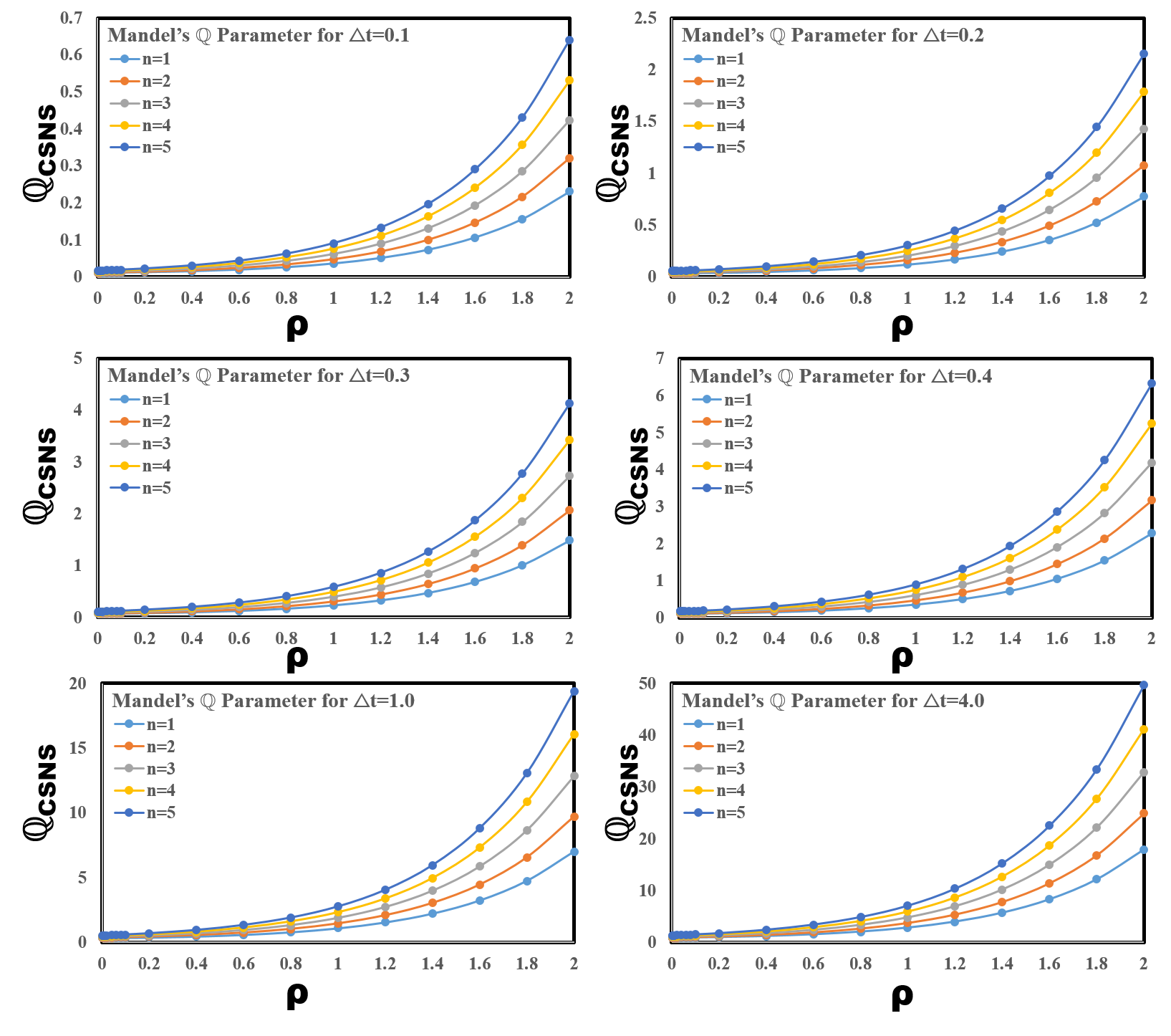}
\caption{Dependence of Mandel's $\mathbf{Q_{\text{CSNS}}}$ Parameter on the squeezing parameter ($\rho $) and quantum number (n). The plot illustrates the relationship highlighting the non-classical characteristics influenced by varying levels of squeezing and quantum states.}
\label{fig:figure_8}
\end{figure}

\begin{figure}[h]%
\centering
\includegraphics[width=0.9\textwidth]{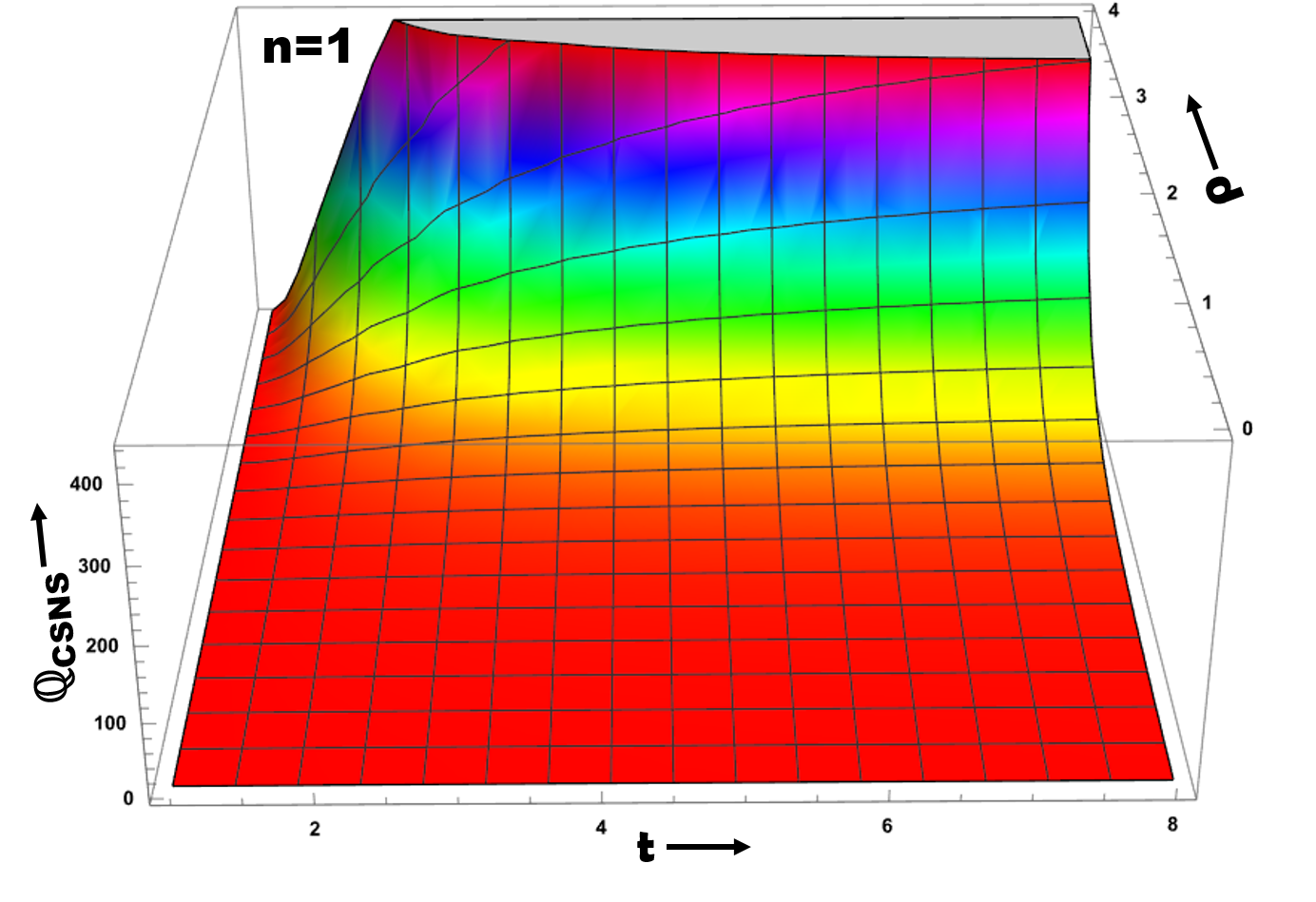}
\caption{3D representation of the variation in Mandel's $\mathbf{Q_{\text{CSNS}}}$ as a function of the squeezing parameter ($\rho$) and time ($t$) for the quantum state with $n=1$, illustrating the dynamic behavior of non-classical states under varying squeezing and temporal evolution.}
\label{fig:figure_9}
\end{figure}

\begin{figure}[h]%
\centering
\includegraphics[width=0.9\textwidth]{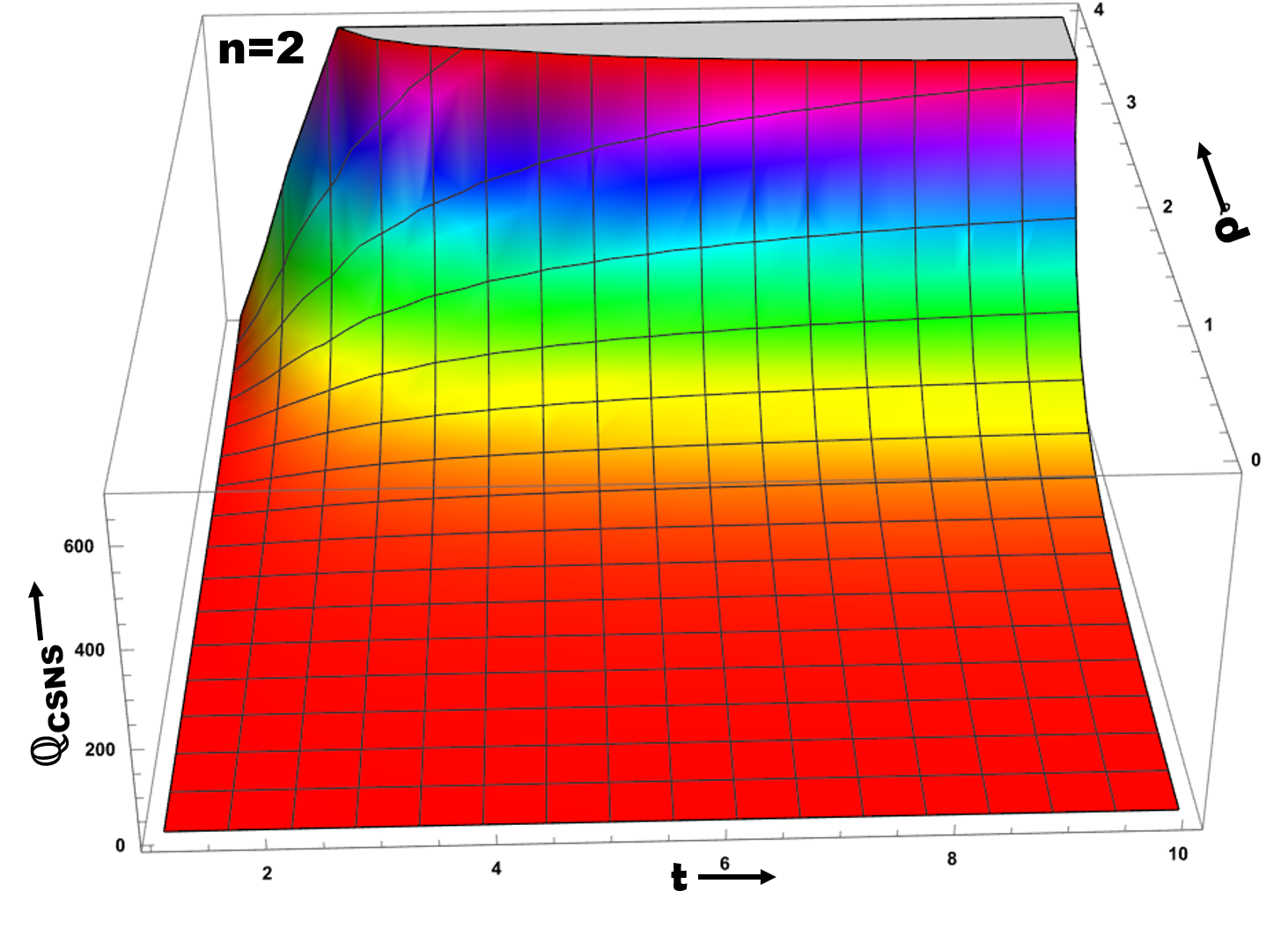}
\caption{3-D representation of the variation in Mandel's $\mathbf{Q_{\text{CSNS}}}$ as a function of the squeezing parameter ($\rho $) and time (t) for the quantum state characterized by n=2. The plot highlights the dynamic behavior of non-classical states under varying squeezing conditions and temporal evolution.}
\label{fig:figure_10}
\end{figure}

\begin{figure}[h]%
\centering
\includegraphics[width=0.9\textwidth]{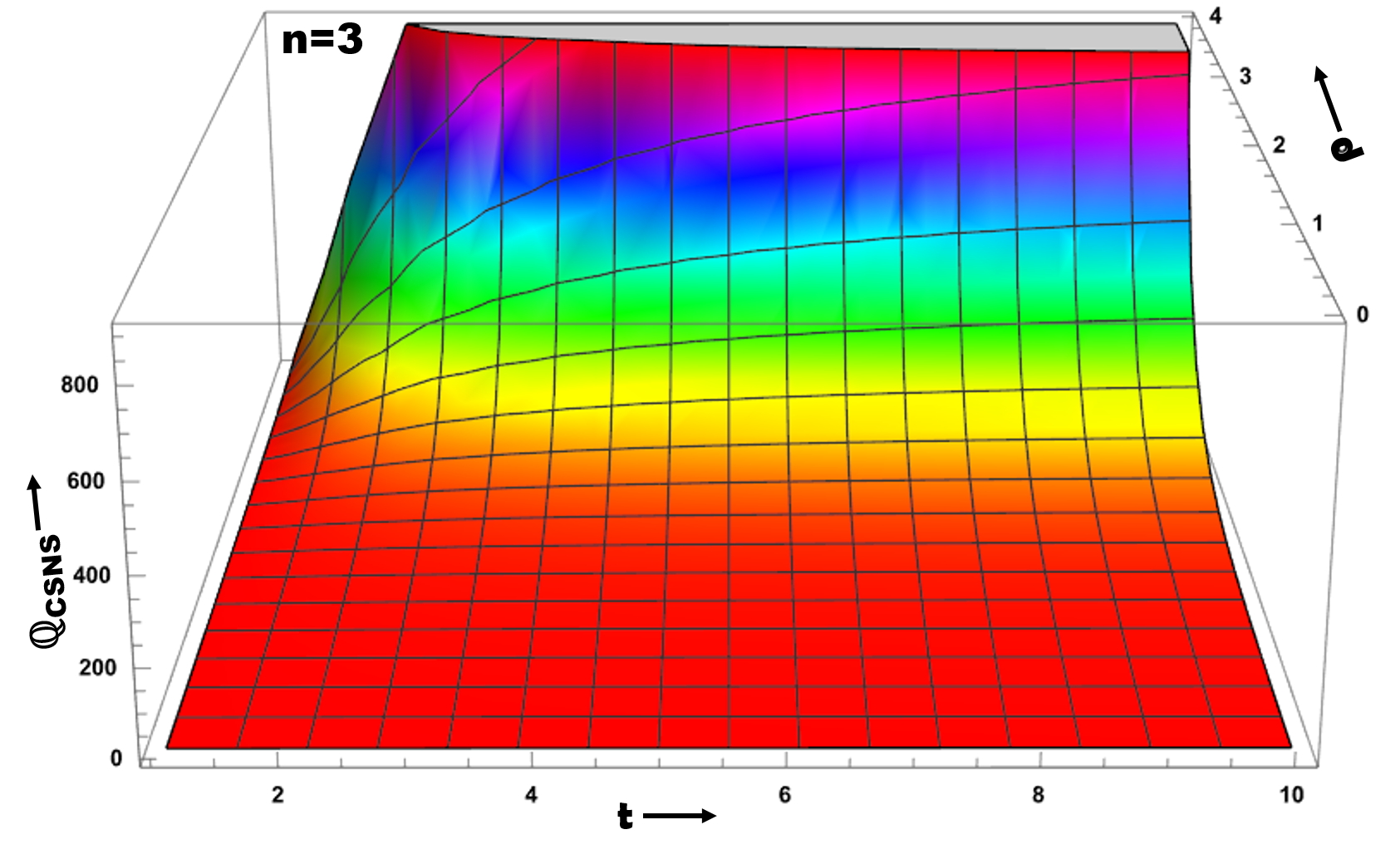}
\caption{3-D representation of the variation in Mandel's $\mathbf{Q_{\text{CSNS}}}$ as a function of the squeezing parameter ($\rho $) and time (t) for the quantum state characterized by n=3. The plot highlights the dynamic behavior of non-classical states under varying squeezing conditions and temporal evolution.}
\label{fig:figure_11}
\end{figure}

\begin{figure}[h]%
\centering
\includegraphics[width=0.9\textwidth]{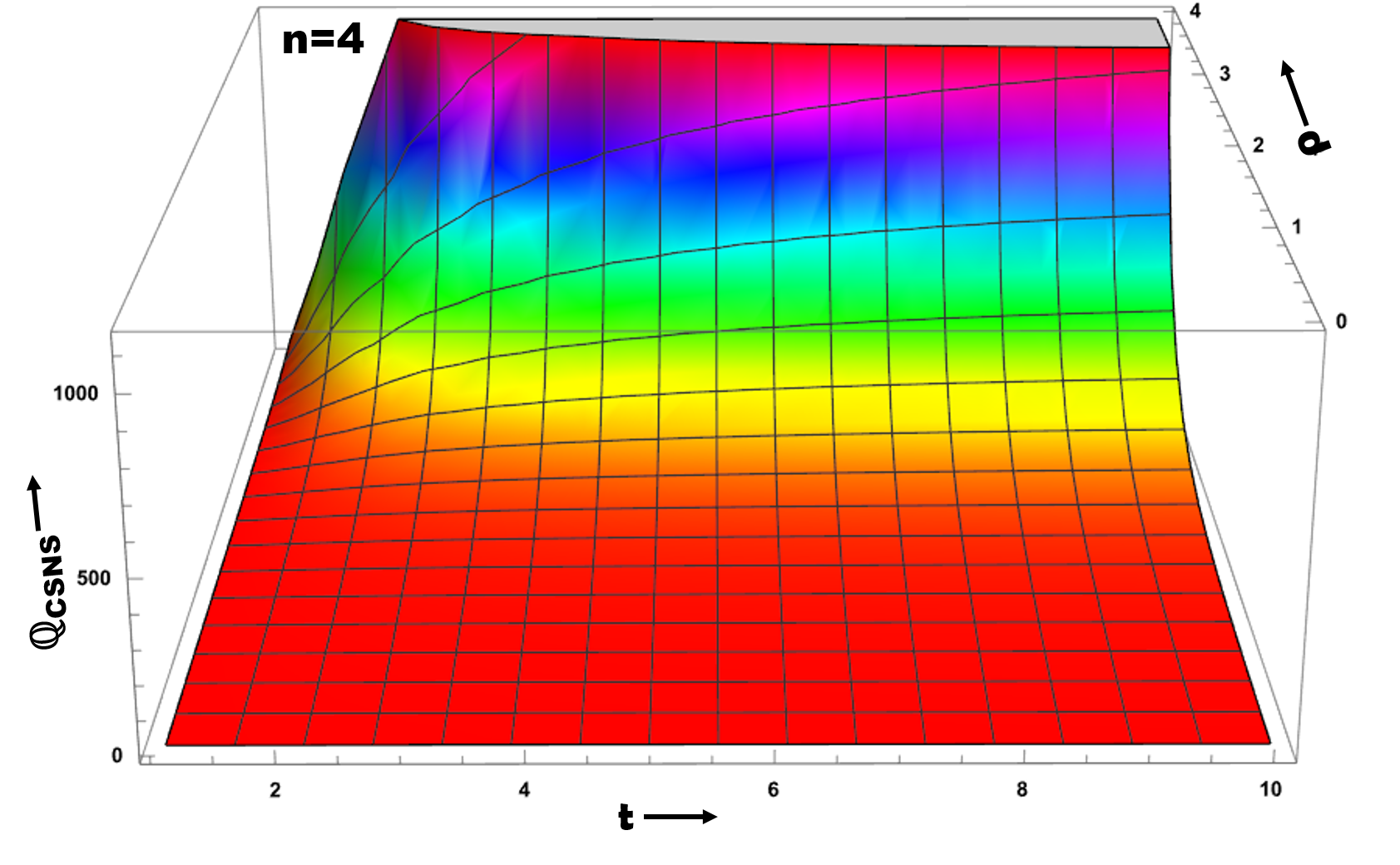}
\caption{3-D representation of the variation in Mandel's $\mathbf{Q_{\text{CSNS}}}$ as a function of the squeezing parameter ($\rho $) and time (t) for the quantum state characterized by n=4. The plot highlights the dynamic behavior of non-classical states under varying squeezing conditions and temporal evolution.}
\label{fig:figure_12}
\end{figure}

\begin{figure}[h]%
\centering
\includegraphics[width=0.9\textwidth]{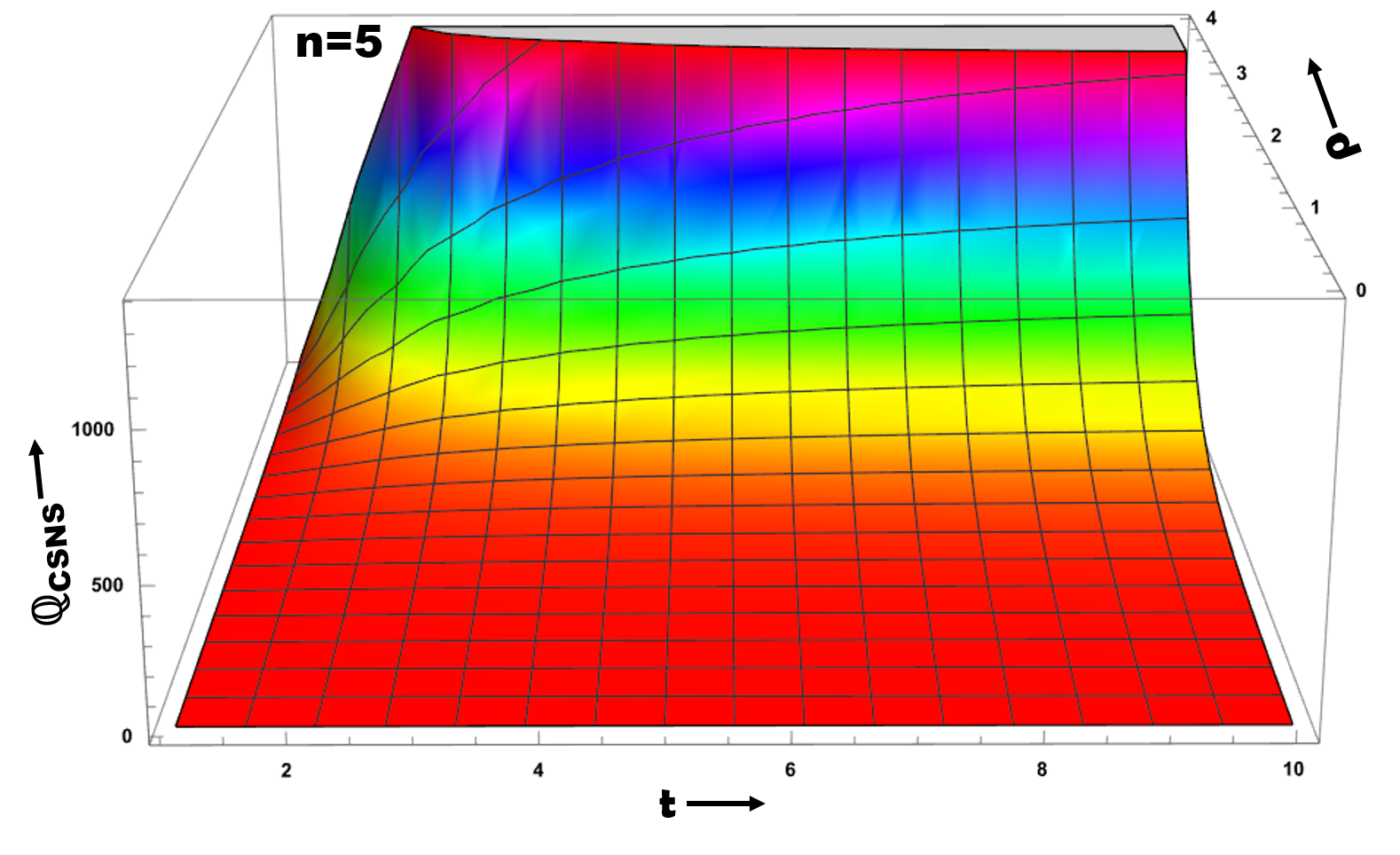}
\caption{3-D representation of the variation in Mandel's $\mathbf{Q_{\text{CSNS}}}$ as a function of the squeezing parameter ($\rho $) and time (t) for the quantum state characterized by n=5. The plot highlights the dynamic behavior of non-classical states under varying squeezing conditions and temporal evolution.}
\label{fig:figure_13}
\end{figure}

\begin{figure}[h]%
\centering
\includegraphics[width=0.9\textwidth]{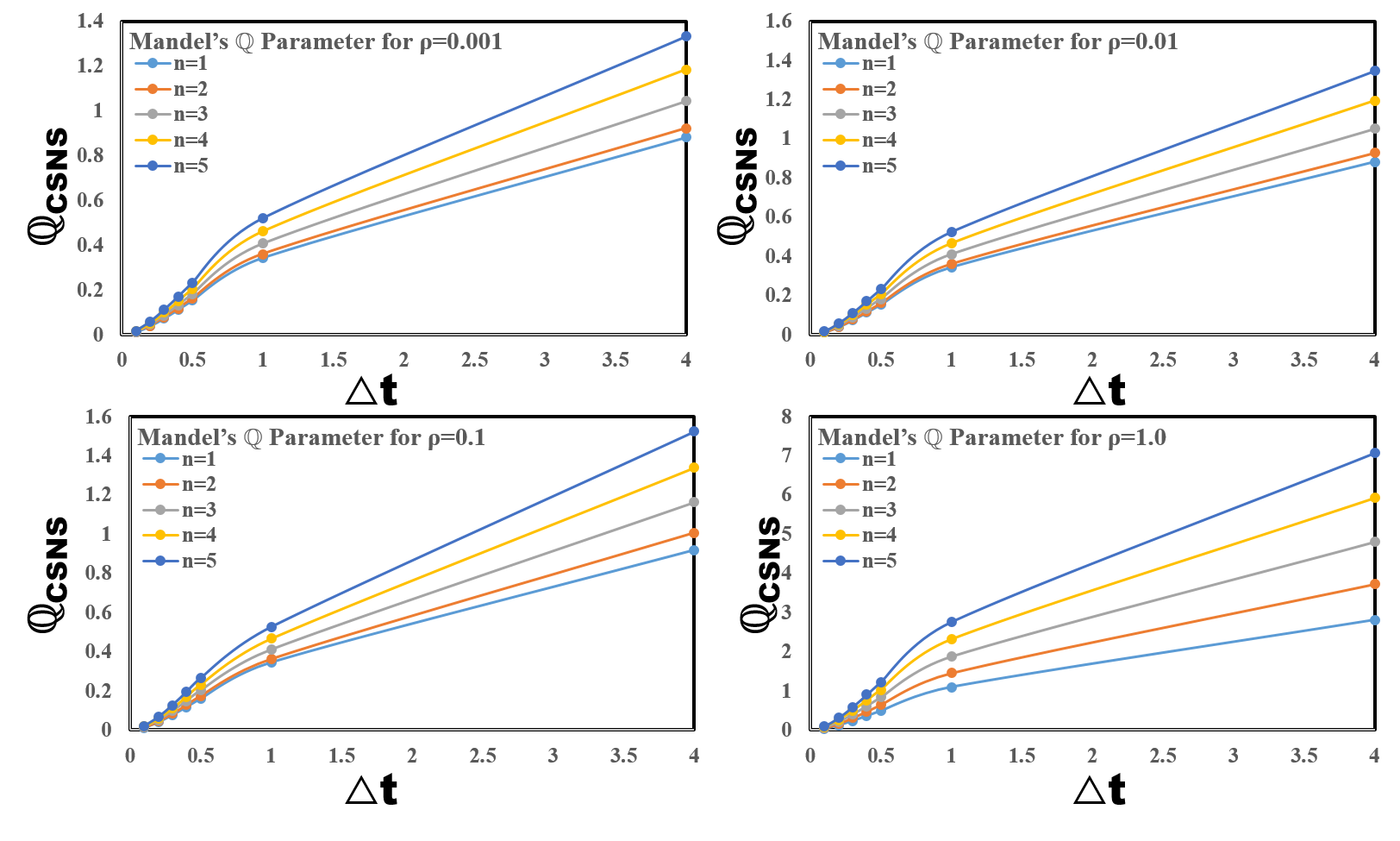}
\caption{Dependence of Mandel's $\mathbf{Q_{\text{CSNS}}}$ Parameter on Time ($t$) and Quantum Number ($n$): The plot highlights the non-classical characteristics as influenced by temporal evolution and quantum states.}
\label{fig:figure_14}
\end{figure}

Hence, Cosmological Mandel's $\mathbb{Q}_{\text{CSNS}}$ Parameter for using quantum mechanical number state for Coherent Squeezed Number State (CSNS) is shown by (\ref {5.2.8}). The $\mathbb{Q}_{\text{CSNS}}$ is the proportional function of squeezing parameter ($\rho $), Coherent state parameter ($\varUpsilon $*, $\varUpsilon $), number state (n) and inversely proportional to various powers of $\mathbf{t} and \mathbf{t}_0$. The dependency of $\mathbb{Q}_{\text{CSNS}}$ is more prominent on the Coherent state parameter ($\varUpsilon $*, $\varUpsilon $) and number state (n) in compression of ($\rho $). Using Eqs. (\ref {5.2.8}), calculated values of $\mathbb{Q}_{\text{CSNS}}$ for various squeezed number state for n=1, 2, 3, 4 is tabulated in tables \ref{tab:table_6}-\ref{tab:table_10}, for simplicity of evaluation here we are assuming $\varUpsilon$*=$\varUpsilon$=m=$\mathbf{t}_0$=1. Numerical values are calculated for the squeezing parameter ($\rho $) ranging between 0.002 to 2.000, while time period $\mathbf{t}$ varies from 1.1 to 6 w.r.t to initial time $\mathbf{t}_0$=1, i.e. the $\triangle\mathbf{t}$ is ranging between 0.1 to 5. Calculations show, for $\rho$ ranging between 0.002 to 1.000 variation in Cosmological Mandel's $\mathbb{Q}_{\text{CSNS}}$ Parameter is much smaller than that for $\rho$ ranging between 1.000 to 2.000 for all number state and $\triangle\mathbf{t}$ taken into consideration. All the values of Cosmological Mandel's $\mathbb{Q}_{\text{CSNS}}$ are positive shows the super-Poissonian non-classical nature of inflaton for all these states. In Eqs. (\ref {5.2.8}) while considering $\varUpsilon$*=$\varUpsilon$=n=0, it converts into eq. for Cosmological Mandel's $\mathbb{Q}$ parameter for squeezed vacuum state \cite{ venkataratnam_density_2008, venkataratnam_oscillatory_2010, venkataratnam_behavior_2013} with the nature of evaluation changes from super-Poissonian non-classical nature to sub-Poissonian non-classical nature. We have also plotted variation of Cosmological Mandel's $\mathbb{Q}$ parameter for various squeezed number states with squeezing parameter $\rho $ in Fig. \ref{fig:figure_8}, that shows increasing nature with increasing $\rho $ as well as n. Fig.  (\ref{fig:figure_9}-\ref{fig:figure_13}) shows a 3-D plot between Cosmological Mandel's $\mathbb{Q}$ parameter, $\rho $, and $\mathbf{t}$ for various number state parameters. While \ref{fig:figure_14} illustrates the dependence of the $\mathbb{Q}_{\text{SNS}}$ parameter on time and quantum state number for a flat FRW universe, highlighting the nonclassical behavior of the inflaton field. The parameter's sensitivity to quantum states and temporal evolution reflects its utility in characterizing statistical properties of the inflaton within semiclassical gravity frameworks.

\section{Results and Discussion}

In this paper, We worked out for consideration the quantum effect on the Squeezed and Coherent State of the flat FRW universe using SCTG. The inclusion of quantum number state in Squeezed and Coherent State is essential for understanding the non-classical nature of gravity, particularly in scenarios involving Particle Production, Density Fluctuations, and Quantum Fluctuations \cite{anderson_effects_1983, campos_semiclassical_1994}. Many universal phenomena can be explained using semiclassical procedures in the lack of a good quantum theory of gravity, where the matter field uses the quantum effect and the background metric remains classical \cite{anderson_effects_1983, campos_semiclassical_1994}. \\
In this context, we analyze the non-classical behavior of SNS and CSNS for a flat FRW universe, with particular emphasis on the cosmological background. Initially, we introduce the quantum number state evolution effect for a massive inflaton in SNS and CSNS using SCTG. Further, we have computed the expression for Cosmological Mandel's $\mathbb{Q}$ Parameter for SNS (\ref {5.1.11}) and CSNS (\ref {5.2.8}).
Our results reveal that the Cosmological Mandel's $\mathbb{Q}$ parameter for CSNS varies with parameters such as the number state ($n$), the coherent state parameter ($\varUpsilon^*, \varUpsilon$), the squeezing parameter ($\rho$), and is inversely proportional to various powers of $\mathbf{t}$ and $\mathbf{t}_0$. In contrast, the $\mathbb{Q}$ parameter for SNS depends on the number state ($n$), the squeezing parameter ($\rho$), and is also inversely proportional to different powers of $\mathbf{t}$ and $\mathbf{t}_0$.
Using suitable approximations in Eqs. (\ref{5.1.11}) and (\ref{5.2.8}), the Cosmological Mandel's $\mathbb{Q}$ parameter for the universe in a vacuum state can be obtained \cite{venkataratnam_density_2008, venkataratnam_behavior_2013, venkataratnam_nonclassical_2010, lachieze-rey_cosmological_1999, takahashi_thermo_1996, xu_quantum_2007}. This demonstrates the validity of the formulation employed in our study using SCTG and SCEE, where the inflaton is minimally coupled with the FRW universe. Here, We investigated the dynamics of a massive inflaton field in an isotropic and homogeneous expanding universe with similar power law growth by considering minimum coupling between a massive inflaton and gravity. In order to examine the behavior of a massive inflaton field within the framework of an isotropic and homogeneous expanding universe with comparable power law expansion, we have taken into consideration the smallest coupling between a massive inflaton and gravity. \\ 
The calculated values of the Cosmological Mandel’s $\mathbb{Q}$  Parameter for various 
n=1,2,3,4,5, using Eqs. (\ref{5.1.11}) and (\ref{5.2.8}), are presented in Tables \ref{tab:table_1}-\ref{tab:table_10}. 2-D plots in Figures (\ref{fig:figure_1}, \ref{fig:figure_7}, \ref{fig:figure_8}, \ref{fig:figure_14}) and 3-D plots in Figures (\ref{fig:figure_2}-\ref{fig:figure_6}, \ref{fig:figure_8}-\ref{fig:figure_13}) provide a clearer visualization of Mandel’s 
$\mathbb{Q}$ Parameter with respect to various variables.

Normally the results demonstrate an increase in $\mathbb{Q}$ Parameter with increasing $\rho $, n as well as $\mathbf{t}$ due to the quantum behavior. The aforementioned analysis shows that all the values of Cosmological Mandel's $\mathbb{Q}$ are positive showing the super-Poissonian non-classical nature of inflaton for all these states. This effectively illustrates the quantum behavior of number states of consideration for SNS and CSNS in the system of a massive inflaton. Mandel's $\mathbb{Q}$ Parameter for both of the above states depends on nature states and $\mathbf{t}$ shows the physical significance of the above analysis to demonstrate quantum properties of oscillating inflaton field concerning general states. So,  In order to better understand the early FRW cosmos, it is important to consider the super-Poissonian non-classical behavior of SNS and CSNS. \\

\bibliography{sn-bibliography.bib}



\end{document}